\documentclass[%
 preprint, 
 amsmath,amssymb,
 aps, physrev,
]{revtex4-2}

\usepackage{graphicx}
\usepackage{dcolumn}
\usepackage{bm}
\usepackage{caption}
\usepackage{subcaption}
\usepackage{dashrule}
\usepackage{xcolor}

\begin{document}

\preprint{APS/123-QED}

\title{\textbf{Bayesian-Enhanced Galerkin-Based Reduced Order Modelling for Unsteady Compressible Flows.} 
}%

\author{Bijie Yang}
 \altaffiliation[Also at ]{Mechanical Engineering, Imperial College London, Londonm UK.}
\author{Chengyuan Liu}%
\author{Lu Tian}%
\affiliation{%
 Aeronautical and Automotive Engineering, Loughborough University, Leicestershire, UK.
}%

\author{Yuping Qian}
\affiliation{
 Automotive Engineering, Tsinghua University, Beijing, China.
}%

\author{Mingyang Yang}
\email{Contact author: myy15@sjtu.edu.cn}
\affiliation{%
 Mechanical Engineering, Shanghai Jiaotong University, Shanghai, China.
}%


\date{\today}

\begin{abstract}

This work proposes a statistical enhanced framework to address the instability and limited predictive capability of conventional Galerkin–Proper Orthogonal Decomposition (Galerkin–POD) models. The method reformulates the correction of the Galerkin-projected ODE system as a statistical inverse problem, in which the coefficients are inferred through Bayesian inference. By accounting for model uncertainty arising from POD mode truncation and data uncertainty introduced by data noise and numerical postprocessing, the framework systematically updates the ODE system coefficients using an analytical, sampling-free solution based on Gaussian likelihood and inverse-Gamma priors. The approach is first validated using a self-sustained oscillating flow over a dimpled surface at a moderate Reynolds number ($Re \approx 3000$), demonstrating stable and accurate reproduction of the temporal dynamics and phase trajectories of coherent structures when compared with direct numerical simulation (DNS). It is then applied to a centrifugal compressor featuring strong tip-leakage vortex breakdown and impeller–diffuser interactions at $Re > 10^5$, where the model successfully captures dominant unsteady structures and frequency characteristics despite limited mode retention. Overall, the results show that Bayesian inference substantially enhances the robustness, stability, and predictive fidelity of Galerkin–POD models for compressible flow systems. The proposed methodology combines the physical interpretability of Galerkin projection with the statistical rigour of Bayesian inference, offering a general, computationally efficient, and uncertainty-aware reduced-order modelling framework for complex fluid dynamic applications.

. 
\end{abstract}

\maketitle


\section{\label{sec:level1}Introduction\protect}

Galerkin–POD is a physics-based, data-driven reduced-order modelling (ROM) that simplifies a system of partial differential equations into a set of ordinary differential equations (ODEs) \cite{holmes2010turbulence}. Using a collection of snapshots obtained from experiments or high-fidelity simulations, Proper Orthogonal Decomposition (POD) extracts an optimal set of modes that capture the maximum kinetic energy of the flow. By projecting the Navier–Stokes equations onto this reduced basis, the Galerkin projection yields a low-dimensional ODE system that governs the temporal evolution of the POD-mode coefficients. This reduction enables efficient approximation of the full flow dynamics at very limited computational cost. Hence, the Galerkin–POD framework has been widely employed to investigate coherent structure dynamics \cite{aubry1988dynamics,rempfer1996investigations,rempfer2003low,rajaee1994low}, analyze flow instabilities \cite{cazemier1998proper}, and develop flow control strategies \cite{ravindran1999proper,prabhu2001influence,bergmann2005optimal}.

In spite of these advantages, the low dimensionality of the POD–Galerkin ROM often leads to a loss of stability and accuracy. This issue becomes more pronounced in fluid dynamics due to the strongly nonlinear nature of the Navier–Stokes equations and the resulting reduced ODE system \cite{holmes2010turbulence}. The nonlinear ODE system is highly sensitive to ODE parameters. Even if a system is correctly initialised, POD–Galerkin models often diverge or converge to erroneous states after long-time integration \cite{sirisup2004spectral}. This challenge is increased in turbulent flows \cite{fick2018stabilized}. As the Reynolds number increases, the energy spectrum of the flow becomes more continuous than sparse. Truncating the POD basis in such a case neglects not only the considerable energy contained in smaller-scale vortices, but also the dissipative effects that naturally stabilise the system. Consequently, most Galerkin–POD studies have been restricted to low-to-moderate Reynolds numbers or flows in simple geometries.

To overcome these challenges, a variety of physics-based strategies have been proposed to stabilise the reduced ODE system. The intermodal energy transfer in POD models bears resemblance to the turbulent kinetic energy cascade \cite{couplet2003intermodal}, suggesting that stabilisation can be achieved by enhancing viscosity or its dissipative effects. Approaches along this line include the introduction of eddy-viscosity transport models such as the Heisenberg model \cite{podvin1998low} or Smagorinsky-type POD closure models \cite{ullmann2010pod,wang2011two}, as well as spectral vanishing viscosity applied through a convolution kernel acting on the higher modes \cite{sirisup2004spectral,tadmor1989convergence}. Another option is to employ the H1 inner product rather than the standard L2 inner product, since the former introduces additional dissipation into the ROM \cite{iollo2000stability}. Beyond viscosity augmentation, stabilisation can also be pursued by modifying the POD basis or the Galerkin projection itself. For instance, the velocity POD basis can be augmented with an extended pressure POD basis to improve stability in incompressible flows \cite{bergmann2009}, while constrained Galerkin projection incorporates long-term attractor properties to enhance robustness \cite{fick2018stabilized}.

In addition to physics-based strategies, the original snapshot data used to construct the POD basis can also be exploited to directly calibrate the coefficients of the reduced ODE system. This calibration can be formulated either as a classic or a statistical inverse problem, where the coefficients are identified by minimising the discrepancy between ROM predictions and the original data \cite{couplet2005calibrated,stabile2020bayesian}. Alternatively, machine learning approaches, or data-driven quadratic ansatz models have been proposed to approximate the coupling effects of truncated POD modes on the resolved dynamics \cite{duriez2017machine,xie2018data}.

Apart from stability issues in Galerkin–POD, another important limitation is that most studies have focused on incompressible flows, while investigations of compressible flows remain very limited. Compressible flows pose additional challenges because the governing equations are coupled through the thermodynamic state equation. For example, Ref.~\cite{rowley2004model} developed a compressible Galerkin–POD framework based on the isentropic Navier–Stokes equations to study two-dimensional self-oscillating cavity flows. In this formulation, density $\rho$ becomes a variable, and compressibility introduces cubic nonlinear terms. To address this, Ref.~\cite{iollo2000stability} proposed reformulating the system in terms of $1/\rho$ rather than $\rho$, which reduces the cubic nonlinearities to quadratic terms. This approach was later applied to investigate three-dimensional self-oscillating cavity flows~\cite{gloerfelt2008compressible} with the calibration method based on regularization inverse problem~\cite{couplet2005calibrated}.

The present study aims to establish a robust, reliable and efficient Galerkin–POD framework for compressible flow systems, capable of accurately representing not only low-Reynolds-number cases but also complex, practically relevant configurations characterised by high Reynolds numbers and intricate geometries. In the framework, the discrepancy between the instable Galerkin–POD and the original high-fidelity unsteady flow field is attributed to both ROM model uncertainty and data uncertainty. Furthermore, Bayesian inference~\cite{kaipio2005statistical} is employed to minimise this discrepancy by correcting the ODE coefficients driven from Galerkin projection. From the framework perspective, model uncertainty arises primarily from POD basis truncation, which neglects the influence of small-scale vortices on the dominant coherent structures. Data uncertainty originates not only from noise in the original measurements but also from additional errors introduced during numerical treatment. For example, dissipation terms (required for Galerkin projection) are often obtained during post-processing of experimental or CFD data, which involves the computation of higher-order derivatives (e.g. gradients of velocity gradients) and thereby amplifies noise. To further improve the generality of the framework and enable its application to complex turbulent flows, where the number of POD modes is significantly larger than in laminar cases, the Bayesian inference is formulated without relying on costly sampling methods. Instead, probability distribution functions are assumed for both the prior (inverse Gamma) and the likelihood (Gaussian), allowing efficient and tractable calibration of the reduced ODE system.

The paper is organised as follows. Section II introduces the detailed methodology of the newly proposed Bayesian-enhanced Galerkin–POD framework for compressible flows. Section III demonstrates the framework on a three-dimensional example of self-oscillating unsteady flow over a dimpled surface, where DNS is used to generate the unsteady data, and provides insights into how Bayesian inference improves ROM reliability. Section IV further evaluates the capability of the framework by applying it to a three-dimensional centrifugal compressor, where LES data are employed to capture complex flow features such as tip-leakage vortex breakdown, boundary-layer separation, wakes, and rotor–stator (impeller-diffuser) interactions. Finally, Section V presents the conclusions.

\section{\label{sec:level2}Bayesian Enhanced Galerk-POD for Compressible Flows}

\subsection{\label{sec:POD} \textbf{Proper orthogonal decomposition}}

Proper Orthogonal Decomposition (POD), also known as the Karhunen–Loève decomposition, Principal Component Analysis (PCA), and Singular System Analysis \cite{kosambi2017statistics, lumey2012stochastic, karhunen1946spektraltheorie, lorenz1956empirical, obukhov1954statistical}, is commonly used to identify an optimal low-dimensional subspace from a dataset. This subspace is considered optimal (or proper) in the sense that it captures the dominant features (or energy) of the data. Moreover, the basis vectors (or modes) of the subspace are orthogonal, facilitating linear reconstruction of the dataset and projection of governing equations onto this subspace.

The theoretical foundation of POD lies in the spectral theory of compact, self-adjoint operators. Let us define a Hilbert space \textbf{\textit{H}} with inner product $(\cdot,\cdot)$. An element $\textbf{f} \in \textbf{\textit{H}}$ represents a possible flow field. The objective of POD is to find a set of optimal basis functions $\boldsymbol{\varphi}$, that maxmize the projetion of $\textbf{f}$ onto $\boldsymbol{\varphi}$:
\begin{equation}
  \mathrm{Max}_{\boldsymbol{\varphi} \in \textbf{\textit{H}}} \frac{ \langle|(\textbf{f},\boldsymbol{\varphi})|^2 \rangle}{\Vert \boldsymbol{\varphi} \Vert^2}
\label{eq:fam}
\end{equation}
$\langle\cdot \rangle$ is the ensemble average, $|\cdot,\cdot|$ is absolute value, and $\Vert \cdot,\cdot \Vert$ is the induced norm. By solving this constrained variational problem, one arrives at the following eigenvalue problem:
\begin{equation}
 \mathcal{R}\boldsymbol{\varphi}=\lambda \boldsymbol{\varphi} 
\label{eq:evp}
\end{equation}
where $\mathcal{R}$ is a linear operator defined as $\mathcal{R}\boldsymbol{\varphi} = \langle(\textbf{f},\boldsymbol{\varphi}) \textbf{f}\rangle$.

In practical applications (e.g., CFD simulations or PIV experiments), the method of snapshots is often used \cite{sirovich1987turbulence}. The Hilbert space is constructed from a collection of time-resolved flow field observations: $\textbf{F}=[\textbf{f}^1,\textbf{f}^2, ...\textbf{f}^M]$, where $\textbf{f}^i=\textbf{f}^i(\textbf{X},t_i)$ denotes the flow field at the time $t_i$ and \textbf{X} irepresents the spatial coordinates of a grid with \textbf{\textit{N}} points. Thus, each $\textbf{f}^i$ is a vector in $\mathbb{R}^N$ and $\textbf{F}$ is an $\textbf{\textit{N}} \times \textbf{\textit{M}}$ matrix. Using the standard inner product $(\mathbf{x},\mathbf{y})=\mathbf{x}^T\cdot \mathbf{y}$, and approximating the ensemble average by the arithmetic mean $\langle \cdot \rangle = \frac{1}{\textbf{\textit{M}}} \sum_{i=1}^{\textbf{\textit{M}}}$, the eigenvalue problem in Eq.~(\ref{eq:evp}) becomes:
\begin{equation}
 \frac{1}{\textbf{\textit{M}}} \textbf{F}\textbf{F}^{T} \boldsymbol{\varphi}^i = \lambda^i \boldsymbol{\varphi}^i
\label{eq:evpsnap}
\end{equation}
Here, $\textbf{F}\textbf{F}^{T}$ is an $\textbf{\textit{N}} \times \textbf{\textit{N}}$ matrix, and $\boldsymbol{\varphi}^i=\boldsymbol{\varphi}^i(\textbf{X})$ the $i^{th}$ patial POD mode. 
The solution of Eq.~(\ref{eq:evpsnap}) is typically obtained via the Singular Value Decomposition (SVD) of the snapshot matrix \textbf{F}:
\begin{equation}
\textbf{F}=\textbf{U}\boldsymbol{\Sigma}\textbf{V}^{T}
\label{eq:svd}
\end{equation}
Here, both $\textbf{U}$ and $\textbf{V}$ are orthogonal matrices ($\textbf{U}^T\textbf{U}=I$, and $\textbf{V}^T\textbf{V}=I$). Particularly, $\textbf{U}_{N \times N}= [\boldsymbol{\varphi}^1,\boldsymbol{\varphi}^2,...\boldsymbol{\varphi}^N]$ contains the normalised spatial POD modes, $\textbf{V}_{M \times M}=[\boldsymbol{c}^1,\boldsymbol{c}^2,...\boldsymbol{c}^M]$ contains the temporal coefficients (right singular vectors), and $\boldsymbol{\Sigma}$ is a rectangular diagonal matrix of singular values:
\begin{equation}
\boldsymbol{\Sigma}=
\begin{bmatrix}
\boldsymbol{\Sigma}_1 \\
0
\end{bmatrix}
\label{eq:sigma}
\end{equation}
where $\boldsymbol{\Sigma}_1 \in \mathbb{R}^{M \times M}$ is a diagonal matrix with entries $\sqrt{\textbf{\textit{M}}\lambda^i}$, arranged in descending order. This decomposition implies that any observed snapshot $\textbf{f}^i$ at time $t_i$ can be represented as a linear combination of the orthonormal basis vectors in $\textbf{U}$, scaled by the corresponding singular values in $\boldsymbol{\Sigma}$, and weighted by the time coefficients in $\textbf{V}$:
\begin{equation}
\textbf{f}^i = \sum_{j=1}^{r} \sqrt{\textbf{\textit{M}}\lambda^j} c^j_i \boldsymbol{\varphi}^j
\label{eq:sigma}
\end{equation}
where r is the rank of $\textbf{F}$. Furthermore, based on the relationship between energy and the autocorrelation of turbulent flows, it can be shown that if $\textbf{F}$ represents turbulent fluctuation fields, then each eigenvalue $\lambda^i$ is twice the average kinetic energy captured by the mode $\boldsymbol{\varphi}^i$ \cite{holmes2010turbulence}. This provides a clear guideline for mode truncation. Normally, only the leading $\textbf{L}$ modes that collectively capture a desired percentage of the total energy are retained. This means the POD basis matrix is truncated as $\boldsymbol{\Phi}=[\boldsymbol{\varphi}^1,\boldsymbol{\varphi}^2,...\boldsymbol{\varphi}^L]$, where $\textbf{L}<<\textbf{N}$.


\subsection{\textbf{Galerkin projection}}

The Galerkin projection reduces a partial differential equation (PDE), such as the compressible Navier–Stokes equations, into a finite set of ordinary differential equations (ODEs) by projecting the governing equations onto the POD basis \cite{Doering1995applied, stone1990study}. A general flow field $\mathbf{f}=\mathbf{f}(\mathbf{X},t)$, not restricted to snapshots at specific times, can be expressed in terms of the first $L$ spatial POD modes with time-dependent coefficients:
\begin{equation}
\mathbf{f}(\mathbf{X},t)= \sum_{i=1}^{L} a^i(t) \boldsymbol{\varphi}^i(\mathbf{X})
\label{eq:rep}
\end{equation}

The evolution of $\mathbf{f}$ is governed by the compressible Navier–Stokes equations, represented by the operator $\mathcal{N}$:
\begin{equation}
\frac{\partial}{\partial t} \mathbf{f}= \mathcal{N}(\mathbf{f})
\label{eq:nsoper}
\end{equation}
To avoid introducing cubic nonlinearities in the reduced system, it is convenient to express the flow field in terms of the primitive variables $\textbf{f}=(\zeta,u,v,w,p)^T$  \cite{iollo2000stability}, where $\zeta=1/\rho$ is the specific volume. Hence, the standard inner product is
\begin{equation}
(\textbf{f}^i,\textbf{f}^j)= \int_{V} (\zeta^i\zeta^j+u^iu^j+v^iv^j+w^iw^j+p^ip^j) dV
\label{eq:standardinnerproduct}
\end{equation}
where $V$ is the flow domain. For the snapshot method, where the flow field information is recorded at $N$ spatial points, the inner product can be simplified as $\sum_{n=1}^{N} (\zeta^i_n \zeta^j_n + u^i_n u^j_n + v^i_n v^j_n + w^i_n w^j_n + p^i_n p^j_n )$, where the subscript $n$ denotes the $n$th grid point.
.Substituting Eq.~(\ref{eq:rep}) into Eq.~(\ref{eq:nsoper}) gives:
\begin{equation}
\sum_{i=1}^{L} \dot{a^i} \boldsymbol{\varphi}^i = \sum_{i=1}^{L}\sum_{j=1}^{L} \mathcal{N}_1(\boldsymbol{\varphi}^i,\boldsymbol{\varphi}^j)a^ia^j+ \sum_{i=1}^{L}\sum_{j=1}^{L} \mathcal{N}_2(\boldsymbol{\varphi}^i,\boldsymbol{\varphi}^j)a^ia^j+ \sum_{i=1}^{L}\mathcal{N}_3(\boldsymbol{\varphi}^i)a^i + \boldsymbol{c}
\label{eq:dens}
\end{equation}
Here, the operator $\mathcal{N}_1$ collects the inviscid contributions of the Navier–Stokes equations (i.e., the Euler terms), while $\mathcal{N}_2$ represents the viscous contributions that provide dissipation or damping in the dynamical system. The operators $\mathcal{N}_3$ and $\mathbf{c}$ arise from the Coriolis and centrifugal forces, respectively, which are essential in a rotating reference frame, such as those encountered in turbomachinery applications. Their explicit forms are:

\begin{equation}
\begin{aligned}
\mathcal{N}_1(\boldsymbol{\varphi}^i,\boldsymbol{\varphi}^j)=
\begin{bmatrix}
-u^i\zeta_x^j -v^i\zeta_y^j -w^i\zeta_z^j + u_x^j\zeta^i + v_y^j\zeta^i + w_z^j\zeta^i\\
-u^iu_x^j -v^iu_y^j -w^iu_z^j - \zeta^i p_x^j \\
-u^iv_x^j -v^iv_y^j -w^iv_z^j - \zeta^i p_y^j \\
-u^iw_x^j -v^iw_y^j -w^iw_z^j - \zeta^i p_y^j \\
-u^ip_x^j -v^ip_y^j -w^ip_z^j -\gamma p^i(u_x^j+v_y^j+w_z^j) \\
\end{bmatrix} 
\\
\mathcal{N}_2(\boldsymbol{\varphi}^i,\boldsymbol{\varphi}^j)=
\begin{bmatrix}
0\\
\mu \zeta^i [u_{xx}^j+u_{yy}^j+u_{zz}^j+\frac{1}{3}(u_{xx}^j+v_{xy}^j+w_{xz}^j)] \\
\mu \zeta^i [v_{xx}^j+v_{yy}^j+v_{zz}^j+\frac{1}{3}(u_{yx}^j+v_{yy}^j+w_{yz}^j)]  \\
\mu \zeta^i [w_{xx}^j+w_{yy}^j+w_{zz}^j+\frac{1}{3}(u_{zx}^j+v_{zy}^j+w_{zz}^j)]  \\
\begin{bmatrix}
\frac{(\gamma -1)k}{R}[\zeta^i (p_{xx}^j+p_{yy}^j+p_{zz}^j)+ p^i(\zeta_{xx}^j+\zeta_{yy}^j+\zeta_{zz}^j)+2(\zeta_{x}^jp_{x}^i+\zeta_{y}^jp_{y}^i+\zeta_{z}^jp_{z}^i)]  \\
(\gamma -1)\mu [2(u_x^iu_x^j+v_y^iv_y^j+w_z^iw_z^j)-2/3(u_x^i+v_y^i+w_z^i)(u_x^j+v_y^j+w_z^j)]\\
(\gamma -1)\mu [(w_y^i+v_z^i)(w_y^j+v_z^j)+(u_z^i+w_x^i)(u_z^j+w_x^j)+(v_x^i+u_y^i)(v_x^j+u_y^j)]
\end{bmatrix}\\
\end{bmatrix}
\\
\mathcal{N}_3(\boldsymbol{\varphi}^i)=
\begin{bmatrix}
0\\
2(\Omega_z v^i-\Omega_y w^i) \\
2(\Omega_x w^i-\Omega_z u^i)  \\
2(\Omega_y u^i-\Omega_x v^i)  \\
0 \\
\end{bmatrix}
, \boldsymbol{c}=
\begin{bmatrix}
0\\
\Omega^2x -\Omega_x(\Omega_x x+\Omega_y y+\Omega_z z)\\
\Omega^2y -\Omega_y(\Omega_x x+\Omega_y y+\Omega_z z)\\
\Omega^2z -\Omega_z(\Omega_x x+\Omega_y y+\Omega_z z)\\
0 \\
\end{bmatrix}
\end{aligned}
\label{eq:nsopermatrix}
\end{equation}
Finally, projecting Eq.~(\ref{eq:dens}) onto $\boldsymbol{\varphi}^k$ and applying the orthogonality of the POD modes yields the reduced-order model:
\begin{equation}
\dot{a}^k =
\sum_{i=1}^{L}\sum_{j=1}^{L} \mathcal{Q}_{1ij}^k a^i a^j +
\sum_{i=1}^{L}\sum_{j=1}^{L} \mathcal{Q}_{2ij}^k a^i a^j +
\sum_{i=1}^{L} \mathcal{L}_{i}^k a^i +
\mathcal{C}^k,
\label{eq:rom}
\end{equation}
with the inner-product operators defined as
\begin{equation}
\mathcal{Q}_{1ij}^k= (\mathcal{N}_1(\boldsymbol{\varphi}^i,\boldsymbol{\varphi}^j),\boldsymbol{\varphi}^k),\quad
\mathcal{Q}_{2ij}^k= (\mathcal{N}_2(\boldsymbol{\varphi}^i,\boldsymbol{\varphi}^j),\boldsymbol{\varphi}^k),\quad
\mathcal{{L}_i}^k= (\mathcal{N}_3(\boldsymbol{\varphi}^i),\boldsymbol{\varphi}^k),\quad
\mathcal{C}^k= (\mathbf{c},\boldsymbol{\varphi}^k).
\label{eq:romops}
\end{equation}
Eq.~(\ref{eq:rom}) provides a dynamical description of the evolution of the POD modes. Once the initial coefficients are determined, the equation can be used to predict their temporal evolution, and through Eq.~(\ref{eq:rep}), reconstruct the corresponding flow field. Moreover, Eq.~(\ref{eq:rom}) shows that the inviscid Euler terms appear as quadratic nonlinear contributions, while the viscous terms, which act to stabilize the system, are also quadratic. The Coriolis force contributes linear terms to the system dynamics, whereas the centrifugal force enters as a constant term, which may correspond to either damping or growth depending on the specific POD modes.

\subsection{\textbf{Bayesian inference}}

The reduction from the full Navier–Stokes equations (Eq.~(\ref{eq:nsoper})) to the Galerkin–POD ODE system (Eq.~(\ref{eq:rom})) inevitably introduces both model uncertainty and data uncertainty. Model uncertainty stems primarily from truncating the POD basis, which neglects the influence of low-energy modes. Although energetically weak, these modes often correspond to small-scale vortices that provide essential damping and stabilisation. Data uncertainty arises not only from measurement noise in the snapshots but also from numerical post-processing. For example, evaluation of the dissipation terms in Eq.~(\ref{eq:nsopermatrix}) requires second-order derivatives, which must be approximated from experimental or CFD data. This procedure can amplify errors, particularly in regions with complex flow phenomena such as vortex shedding or separation. As a result, the reduced system ((\ref{eq:rom})) often suffers from instability and poor predictive capability.

To address these challenges, we formulate the calibration of the reduced system as a statistical inverse problem using Bayesian inference. The aim is to update the coefficients in Eq.~(\ref{eq:romops}) so that the predicted dynamics of Eq.~(\ref{eq:rom}) are consistent with the observed snapshot data.

Taking the k-th mode as an example, the temporal coefficients are obtained from POD reconstruction Eq.~(\ref{eq:sigma}) at \textbf{\textit{M}} discrete time steps:
\begin{equation}
\mathbf{a}^k=(a^k_1,a^k_2,...,a^k_\mathbf{M})^T \in \mathbb{R}^M
\end{equation}
where $a^k_i=\sqrt{\textbf{\textit{M}}\lambda^j} c^j_i$. The corresponding time derivatives are computed numerically, e.g. via a central difference scheme:
\begin{equation}
\mathbf{y}^k=(\dot{a}^k_1,\dot{a}^k_2,...,\dot{a}^k_\mathbf{M})^T
\end{equation}
For clarity, in Eq.~(\ref{eq:rom}), the quadratic term coefficients $\mathcal{Q}_{1}$ and $\mathcal{Q}_{2}$ and combined into a single coefficient $\mathcal{Q}$, and the terms $a^ia^j$ and $a^ja^i$ are merged for $i \neq j$. The reduced system can then be expressed compactly as: 
\begin{equation}
\mathbf{y}^k = \mathbf{A} \mathbf{x}^k
\label{eq:romreo}
\end{equation}
Here, $\mathbf{A} \in \mathbb{R}^{M\times D}$ is constructed from the observed data (e.g., $a^{i}a^{j}, a^i$), with $\textbf{\textit{D}}=\textbf{\textit{L}}\frac{\textbf{\textit{L}}+1}{2}+\textbf{\textit{L}}+1$. $\mathbf{x}^k\in \mathbb{R}^{P}$ represents the uncertain coefficient vector (containing ${Q}_{ij}^k$, ${L}_{i}^k$, $\mathcal{C}^k$) with prior information from Eq.~(\ref{eq:rom}).

Bayesian inference is based on Bayes' Theorem. It estimates $\mathbf{x}$ by treating it as a random variable and updating its probability distribution based on the observed data $\mathbf{y}$. In practice, observations are corrupted by uncertainty $\epsilon$, giving (for simplicity, the superscript $'k'$ is omitted in the following discussion):
\begin{equation}
\mathbf{y} = \mathbf{A} \mathbf{x} + \epsilon
\label{eq:blromreo}
\end{equation}
Applying Bayes’ theorem:
\begin{equation}
P(\mathbf{x}  | \mathbf{y} ) = \frac {P(\mathbf{y}  | \mathbf{x} )P(\mathbf{x} )}{P(\mathbf{y} )}
\label{eq:bimath}
\end{equation}
Here $P(\mathbf{x}  | \mathbf{y} )$ is the probability density function of $\mathbf{x}$ at the condition of observed $\mathbf{y}$. It is the posterior probability or the inverse problem, which represents the updated belief about the hypothesis $\mathbf{x}$ after observing the data $\mathbf{y}$. $P(\mathbf{y}  | \mathbf{x} )$ is the likelihood or the forward problem. $P(\mathbf{x} )$ is the prior probability, which represents the initial belief about the hypothesis before observing the data. $P(\mathbf{y} )$ is the evidence that normalizes the posterior probability.

If $\epsilon$ is a multivariate normal distribution with mean $0$ and a known covariance matrix $\sigma^2I$, the likelihood is:
\begin{equation}
P(\mathbf{y}  | \mathbf{x} ) =\frac{1}{(2\pi)^{\frac{M}{2}}} \frac{1}{\vert \sigma^2I\vert^{\frac{1}{2}}}exp[-\frac{1}{2}(\mathbf{y}-A\mathbf{x})^T [\sigma^2I]^{-1}(\mathbf{y}-A\mathbf{x})]
\label{eq:bimathlike}
\end{equation}
The prior distribution of the regression coefficients can also be assumed to has a multivariate normal distribution with mean $\mathbf{x}_0$ and covariance matrix $\sigma^2V_0$:
\begin{equation}
P(\mathbf{x}) =\frac{1}{(2\pi)^{\frac{D}{2}}} \frac{1}{\vert \sigma^2V_0\vert^{\frac{1}{2}}}exp[-\frac{1}{2}(\mathbf{x}-\mathbf{x}_0)^T [\sigma^2V_0]^{-1}(\mathbf{x}-\mathbf{x}_0)]
\label{eq:bimathprior1}
\end{equation}
where $V_0 \in \mathbb{R}^{D \times D}$ is a symmetric positive definite matrix. Given the prior and the likelihood show above, the posterior is also a multivariate normal distribution with mean $\mathbf{x}_D$ and covariance matrix $\sigma^2V_D$:
\begin{equation}
P(\mathbf{x}  | \mathbf{y} ) =\frac{1}{(2\pi)^{\frac{D}{2}}} \frac{1}{\vert \sigma^2V_D\vert^{\frac{1}{2}}}exp[-\frac{1}{2}(\mathbf{x}-\mathbf{x}_D)^T [\sigma^2V_D]^{-1}(\mathbf{x}-\mathbf{x}_D)]
\label{eq:bimathpost}
\end{equation}
\begin{equation}
\mathbf{x}_D=(V_0^{-1}+A^TA)^{-1}[V_0^{-1}\mathbf{x}_0+A^T\mathbf{y}]
\label{eq:bimathmean}
\end{equation}
\begin{equation}
V_D=(V_0^{-1}+A^TA)^{-1}
\label{eq:bimathconvar}
\end{equation}
In many practical situations, the variance of the error term $\sigma^2$ is also unknown. This leads naturally to a hierarchical Bayesian inference formulation, where a prior distribution must also be specified for $\sigma^2$. For turbulent flows, the number of modes retained can be very large ($\textbf{\textit{D}}>>10$), making sampling-based approaches impractical. To obtain a tractable analytical solution, $\sigma^2$ is assigned an inverse-Gamma prior distribution with parameters parameters $E$ and $\sigma^2_0$:
\begin{equation}
P(\sigma^2) =\frac{(E\sigma^2_0)^{\frac{E}{2}}}{2^{\frac{E}{2}}\Gamma(\frac{E}{2})} \frac{1}{(\sigma^2)^{\frac{E}{2}+1}}exp(-\frac{1}{2}\frac{E\sigma^2_0}{\sigma^2})
\label{eq:bimathprior2}
\end{equation}
Meanwhile, Eq.~(\ref{eq:bimathlike}) represents the likelihood $P(\mathbf{y}  | \mathbf{x}, \sigma^2 )$, and Eq.~(\ref{eq:bimathprior1}) gives the prior $P(\mathbf{x}  | \sigma^2 )$. Consequently, the posterior distribution of the regression coefficients follows a multivariate Student's distribution:
\begin{equation}
P(\mathbf{x}  | \mathbf{y} ) = \frac{\Gamma
(\frac{M+E+D}{2})}{\Gamma(\frac{M+E}{2})[\pi(M+E)]^{\frac{D}{2}}\vert \sigma^2_DV_D\vert^{\frac{1}{2}}}[1+\frac{1}{M+E}(\mathbf{x}-\mathbf{x}_D)^T [\sigma^2_DV_D]^{-1}(\mathbf{x}-\mathbf{x}_D)]^{-\frac{M+E+D}{2}}
\label{eq:bimath}
\end{equation}
with mean $\mathbf{x}_D$, variance $\frac{M+E}{M+E-2}\sigma^2_DV_D$, where
\begin{equation}
\sigma^2_D=\frac{1}{M+E}[E\sigma^2_0+(\mathbf{y}-A\mathbf{x}_0)^T [AV_0A^T+I]^{-1}(\mathbf{y}-A\mathbf{x}_0)]
\label{eq:bimath}
\end{equation}

In addition, the posterior mean $\mathbf{x}_D$ can be rewritten as:
\begin{equation}
\mathbf{x}_D=(V_0^{-1}+A^TA)^{-1}[V_0^{-1}\mathbf{x}_0+(A^TA)\mathbf{x}_{ols}]
\label{eq:bimathmeanre}
\end{equation}
where $\mathbf{x}_{ols}=(A^TA)^{-1}A^T\mathbf{y}$, which is the ordinary least squares (OLS) estimator of the regression coefficients of Eq.~(\ref{eq:romreo}). Thus, This results shows taht the Bayesian posterior mean $\mathbf{x}_{D}$ is the weighted combination of the prior knowledge $\mathbf{x}_0$ and the OLS estimate derived from the observed data $\mathbf{x}_{ols}$. 





\section{\textbf{Dimpled Surface}}

To demonstrate the capability of the Bayesian inference on Galerkin–POD approach, the methodology described above has been applied to two cases with different Reynolds numbers and complexity: a self-sustained oscillating flow over a single dimpled surface ($Re \sim 10^3$) and a centrifugal compressor exhibiting tip-leakage vortex breakdown and blade–blade interactions ($Re \sim 10^5$).

\subsection{\textbf{Physical problem and CFD results}}
\begin{figure}
    \includegraphics[width=0.5\linewidth]{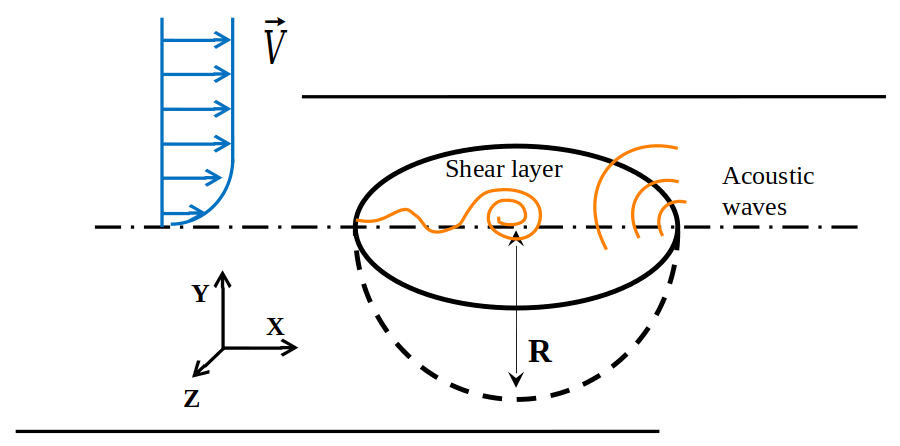}
    \caption{Sketch of the dimpled surface.}
    \label{fig:DSSketch}
\end{figure}

A schematic of the dimpled surface is shown in Fig.~\ref{fig:DSSketch}, where an incoming boundary layer flows over a hemispherical cavity. Self-sustained oscillations are observed: the shear layer originating from the main boundary layer amplifies as it passes over the cavity and then impinges on the downstream edge. The resulting acoustic waves propagate upstream, interacting with the shear layer and strengthening its instability.

The boundary conditions and computational domain follow previous studies of rectangular cavities with comparable flow characteristics ~\cite{rowley2004model, gloerfelt2008compressible}. The free-stream Mach number is $M_{\infty}=0.6$, and the Reynolds number based on the cavity diameter ($D$) is $Re=3000$. All wall boundaries are treated as adiabatic and no-slip. The computational domain is sufficiently large to capture the radiated acoustic field, extending $14D$ in the streamwise direction, $10D$ in the wall-normal direction, and $10D$ in the spanwise direction. A structured O-grid was used to mesh the hemispherical cavity. The mesh was organised such that, on the spanwise symmetric plane of the hemispherical cavity, the mesh quality is comparable to that of two-dimensional reference studies. In particular, on this plane, the grid resolution consists of $903 \times 201$ points above the cavity in the streamwise and wall-normal directions respectively, and $303 \times 152$ points inside the cavity. The mesh was refined near the wall to resolve boundary-layer accurately. The simulations were performed using OpenFOAM’s rhoPimpleFoam solver to resolve the unsteady compressible Navier–Stokes equations. The time step was set to $\Delta t \approx 2\times 10^{-10}$ so that Courant-Friedrichs-Lewy (CFL) number is less than one for any mesh element.

\begin{figure}
    \includegraphics[width=0.8\linewidth]{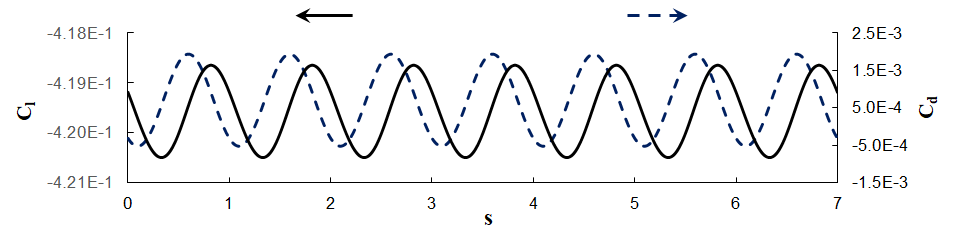}
    \caption{Temporal variation of lift and drag coefficients.}
    \label{fig:DSClCd}
\end{figure}

\begin{figure}
  \begin{center}
    \begin{subfigure}[t]{0.4\textwidth}
      \centering
      \includegraphics[width=0.9\textwidth]{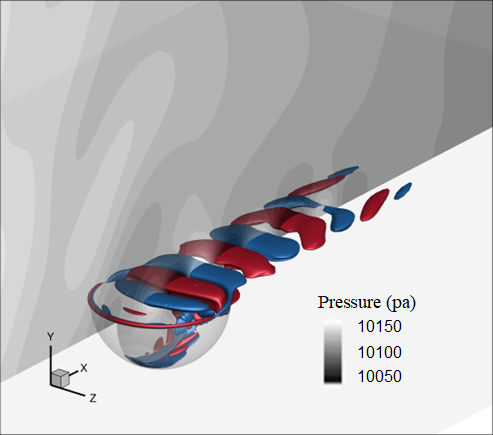}
      \caption{$s=0$}
      \label{fig:CFDFF1}
    \end{subfigure}
    \quad
    \begin{subfigure}[t]{0.4\textwidth}
      \centering
      \includegraphics[width=0.9\textwidth]{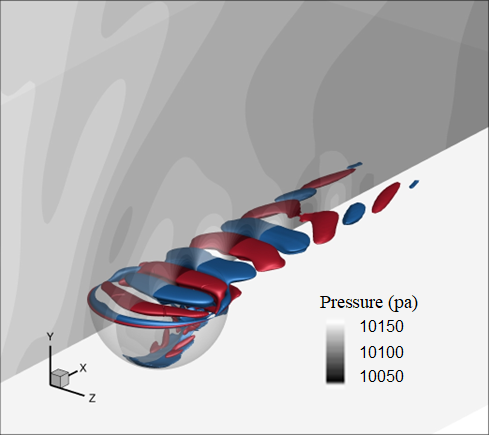}
      \caption{$s=0.25$}
      \label{fig:CFDFF2}
    \end{subfigure}
     \begin{subfigure}[t]{0.4\textwidth}
      \centering
      \includegraphics[width=0.9\textwidth]{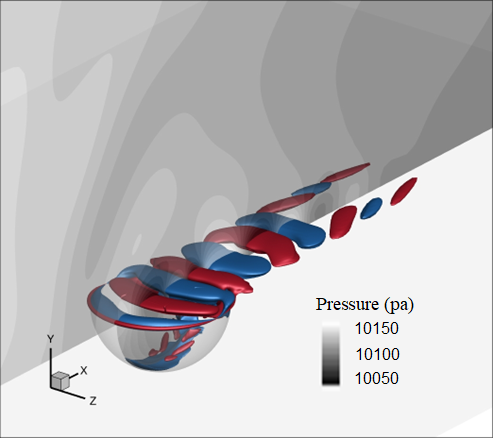}
      \caption{$s=0.5$}
      \label{fig:CFDFF3}
    \end{subfigure}
    \quad
    \begin{subfigure}[t]{0.4\textwidth}
      \centering
      \includegraphics[width=0.9\textwidth]{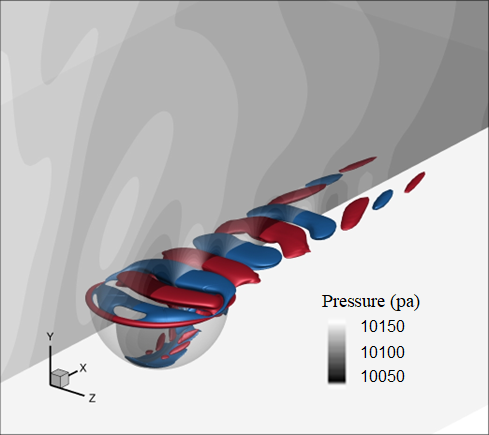}
      \caption{$s=0.75$}
      \label{fig:CFDFF4}
    \end{subfigure}
  \end{center}
  \caption{Flow field evolution at different times within one oscillation period based on Q-criteria and acoustic waves.}
  \label{fig:DSFlowfield}
\end{figure}

Overall 204 CFD results over seven oscillation periods were collected after a physical time of $tD/U=500$, at which point the self-sustained oscillations were fully developed. The instantaneous lift and drag coefficients ($C_l$ and $C_d$) within the seven periods are presented in Fig.~\ref{fig:DSClCd}, showing a dominant harmonic oscillation characterized by a single frequency component $f$. The corresponding Strouhal number of the frequency is $S_t=fD/U_\infty=0.88$, which is $20\%$ higher than a 2D rectangular cavity. Figure~\ref{fig:DSFlowfield} illustrates the evolution of the flow field over one complete oscillation cycle. Iso-surfaces of the Q-criterion is used to visualize vortical structures, and pressure contours on the symmetric spanwise plane is applied to visualize the acoustic waves. The non-dimensional time is defined as $s=f\cdot t$. The blue and red iso-surfaces correspond to equal magnitudes of the Q-criterion with opposite signs, representing a pair of vortices.

\subsection{\textbf{Application of Galerkin-POD}}

The POD analysis was performed by applying singular value decomposition Eq.~(\ref{eq:svd}) to the data matrix constructed from 204 instantaneous CFD snapshots. Prior to the POD, the CFD results were interpolated onto a coarser but uniform mesh. The original computational mesh contains approximately $50 \times 10^6$ elements; thus, a dataset comprising $204$ snapshots would require around $816$ GB of storage, which poses a significant challenge for performing SVD with the available computational resources. Moreover, since the original mesh is highly refined near the wall, while the purpose of POD is to extract the dominant coherent structures across the entire flow domain, the use of a uniform mesh is preferred.

The eigenvalues of the POD modes are presented in Fig.~\ref{fig:DSFlowfield}, with the mean flow subtracted. As discussed in Sec.~\ref{sec:POD}, $\lambda_n$ represents the time-averaged energy associated with the $n^{th}$ OD mode. The results indicate that the first two modes dominate the flow dynamics, capturing approximately $99.5\%$ of the total oscillation energy ($\sum_{i=1}^{2} \lambda_{i} / \sum_{i=1}^{204} \lambda_{i} \approx 0.995$). Beyond these, the modal energy decreases rapidly until about $n\approx 14$, after which the rate of decay slows down. Notably, the first four modes together account for $99.99\%$ of the total energy, and the first six modes together account for more than $99.999\%$.
\begin{figure}
    \includegraphics[width=0.4\linewidth]{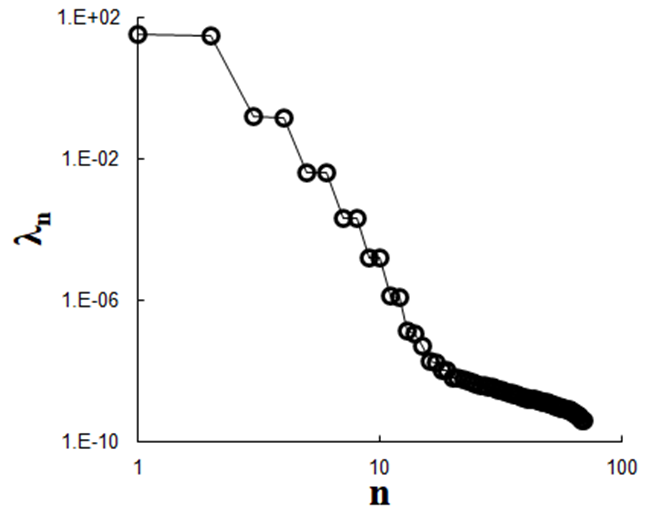}
    \caption{POD eigenvalues $\lambda_n$.}
    \label{DSLamda}
\end{figure}

\begin{figure}
  \begin{center}
    \begin{subfigure}[t]{0.32\textwidth}
      \centering
      \includegraphics[width=0.99\textwidth]{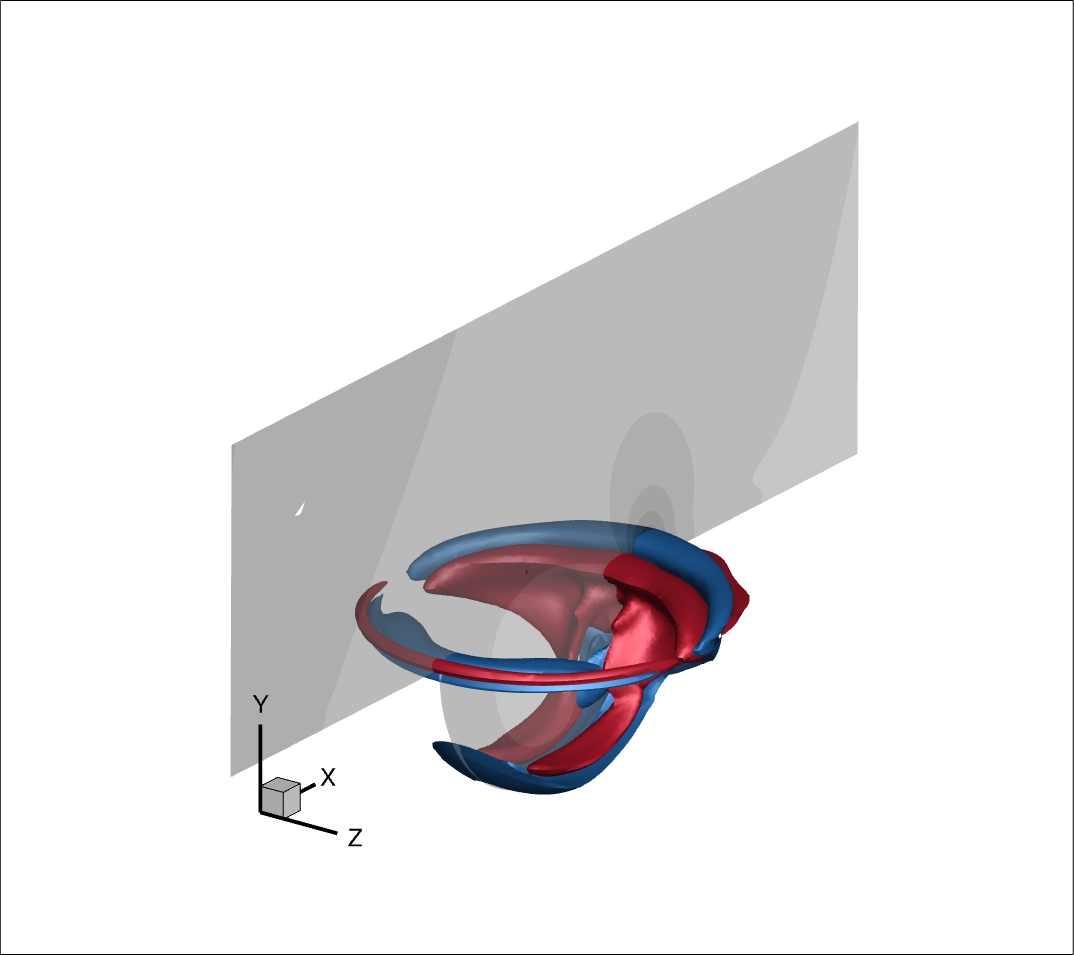}
      \caption{Mean flow}
      \label{fig:POD0}
    \end{subfigure}
    \\
    \begin{subfigure}[t]{0.32\textwidth}
      \centering
      \includegraphics[width=0.99\textwidth]{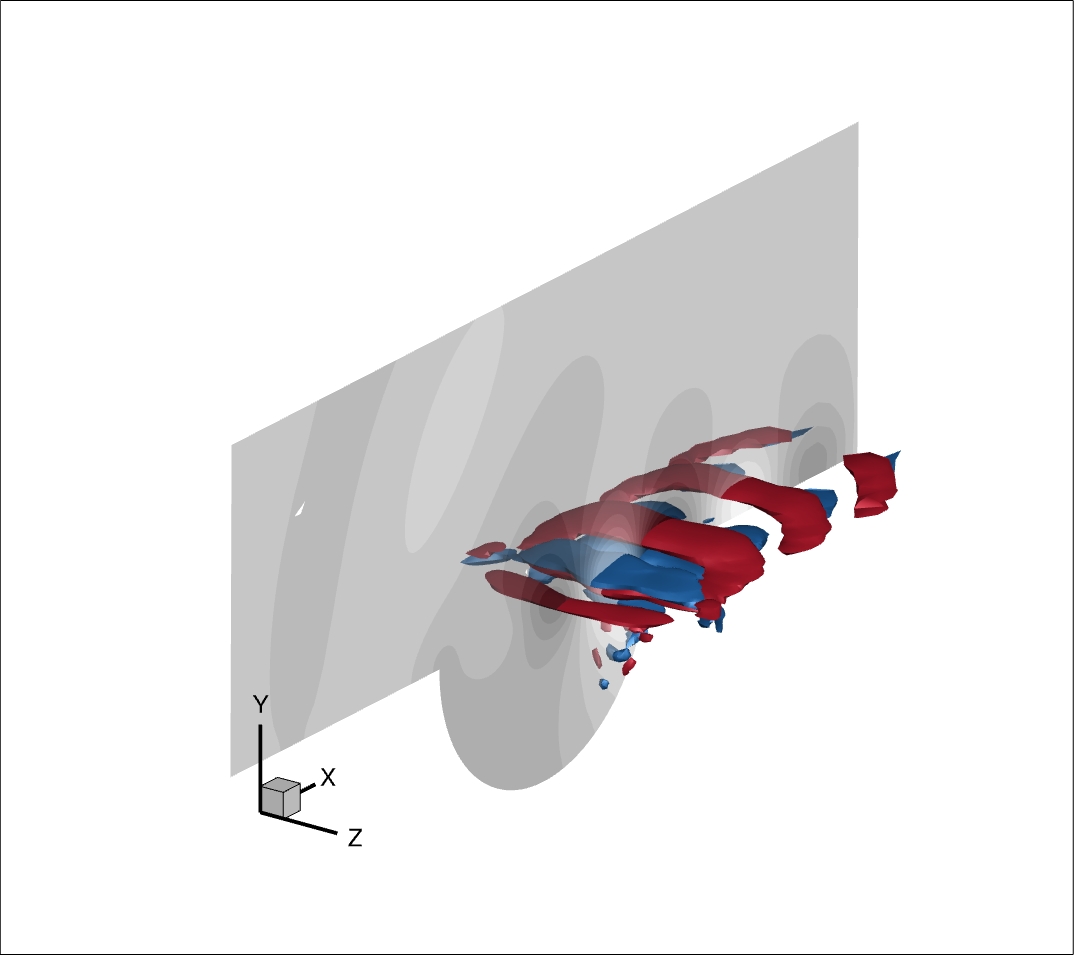}
      \caption{Mode 1}
      \label{fig:POD1}
    \end{subfigure}
    \begin{subfigure}[t]{0.32\textwidth}
      \centering
      \includegraphics[width=0.99\textwidth]{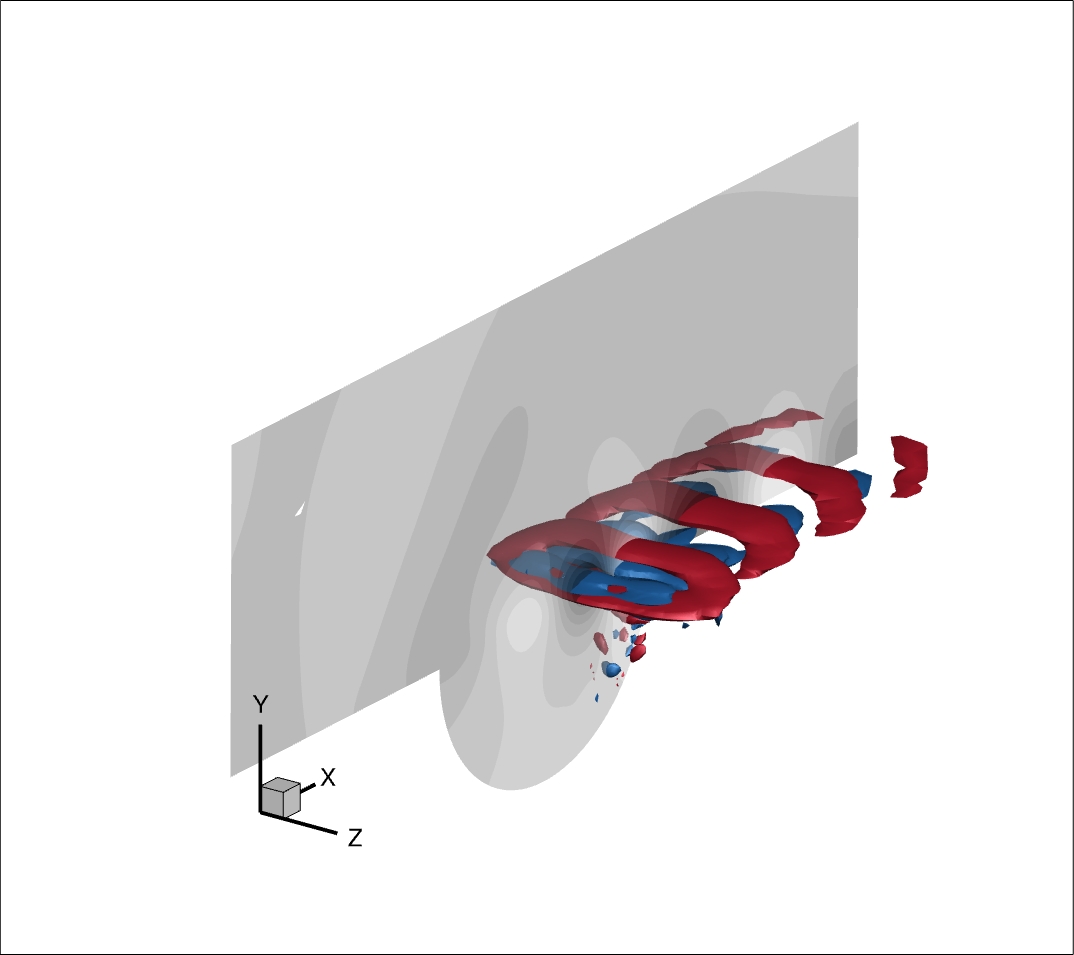}
      \caption{Mode 2}
      \label{fig:POD2}
    \end{subfigure}
     \begin{subfigure}[t]{0.32\textwidth}
      \centering
      \includegraphics[width=0.99\textwidth]{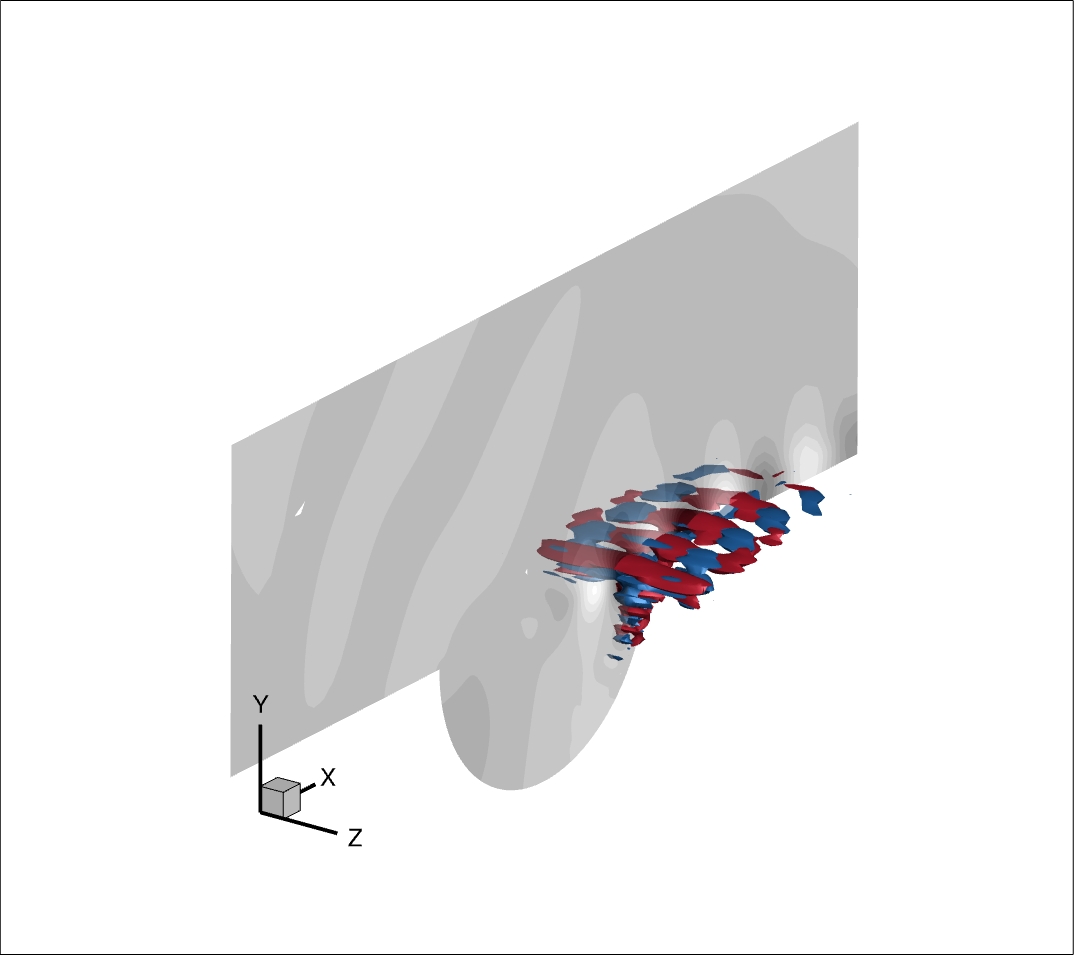}
      \caption{Mode 3}
      \label{fig:POD3}
    \end{subfigure}
    \quad
    \begin{subfigure}[t]{0.32\textwidth}
      \centering
      \includegraphics[width=0.99\textwidth]{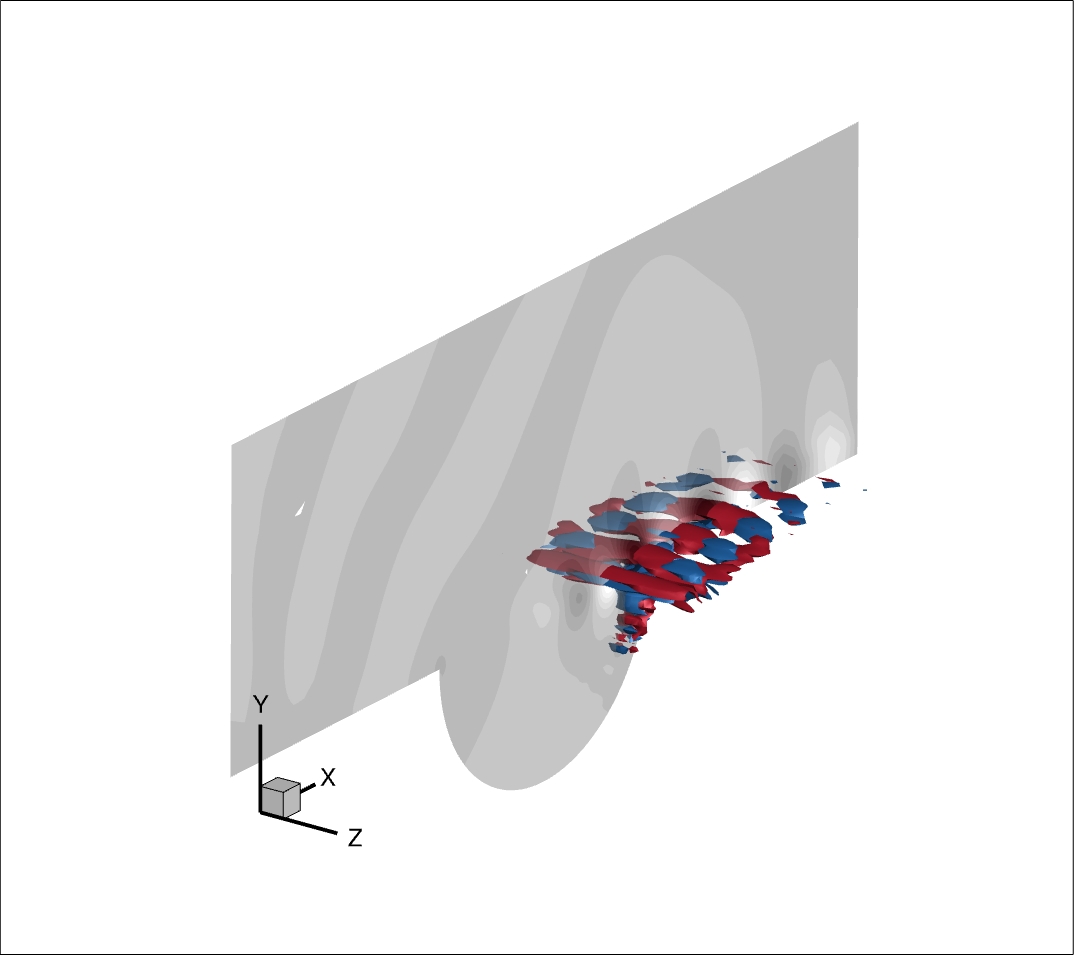}
      \caption{Mode 4}
      \label{fig:POD4}
    \end{subfigure}
     \begin{subfigure}[t]{0.32\textwidth}
      \centering
      \includegraphics[width=0.99\textwidth]{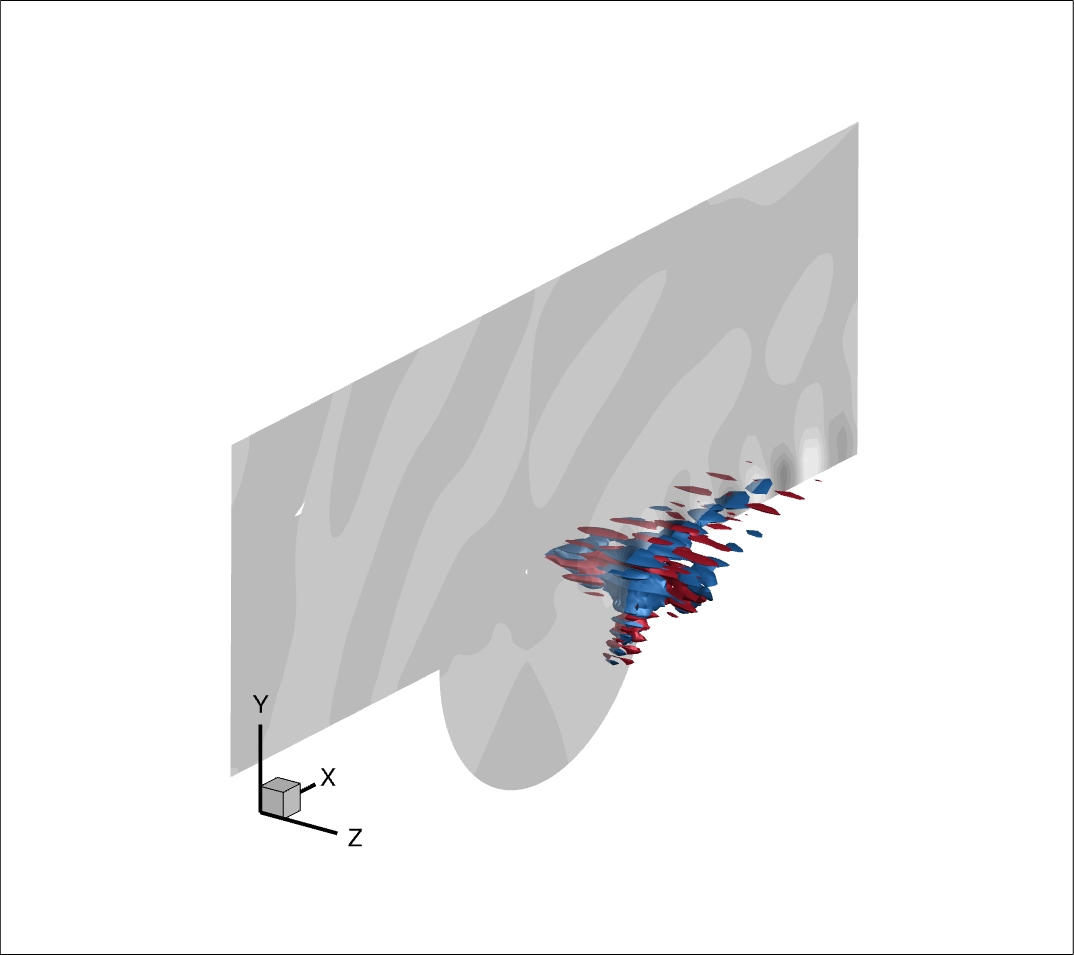}
      \caption{Mode 5}
      \label{fig:POD5}
    \end{subfigure}
     \begin{subfigure}[t]{0.32\textwidth}
      \centering
      \includegraphics[width=0.99\textwidth]{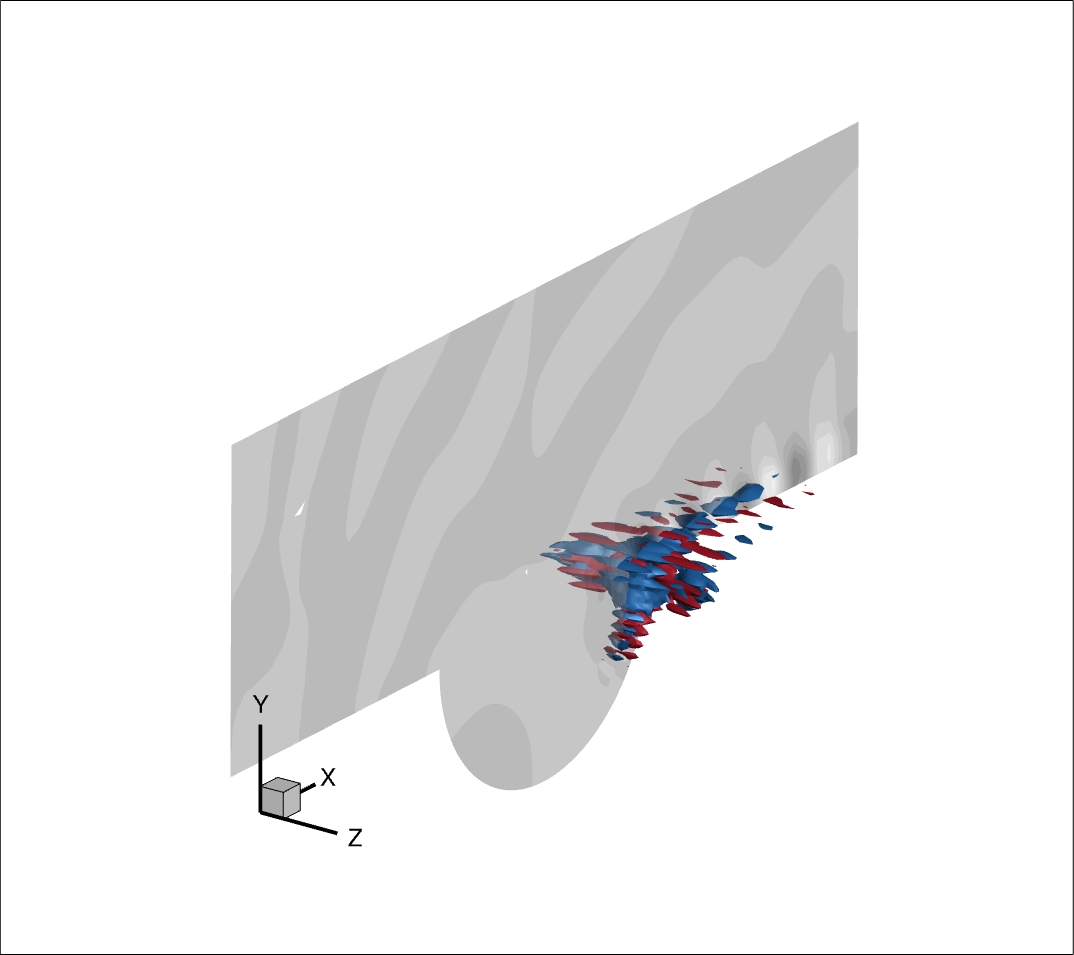}
      \caption{Mode 6}
      \label{fig:POD6}
    \end{subfigure}
  \end{center}
  \caption{POD modes based on Q-criteria and acoustic waves.}
  \label{fig:DSmodes}
\end{figure}

\begin{figure}
  \begin{center}
    \begin{subfigure}[t]{0.7\textwidth}
      \centering
      \includegraphics[width=0.9\textwidth]{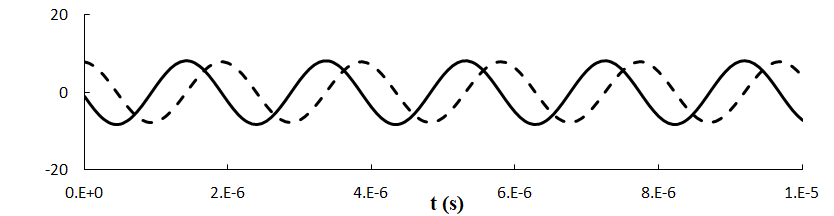}
      \caption{\rule[0.5ex]{0.5cm}{0.5pt} Model 1, \hdashrule[0.5ex]{0.5cm}{0.5pt}{1mm 0.5mm} Model 2}
      \label{fig:DSCFDCoef12}
    \end{subfigure}
    \quad
    \begin{subfigure}[t]{0.7\textwidth}
      \centering
      \includegraphics[width=0.9\textwidth]{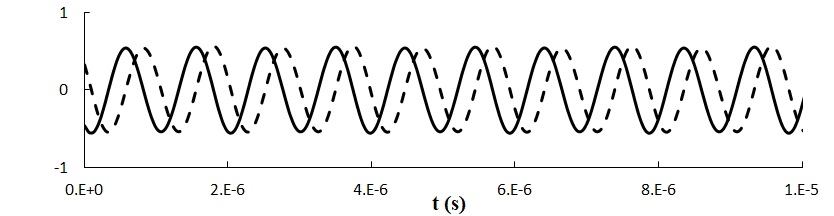}
      \caption{\rule[0.5ex]{0.5cm}{0.5pt} Model 3, \hdashrule[0.5ex]{0.5cm}{0.5pt}{1mm 0.5mm} Model 4}
      \label{fig:DSCFDCoef34}
    \end{subfigure}
     \begin{subfigure}[t]{0.7\textwidth}
      \centering
      \includegraphics[width=0.9\textwidth]{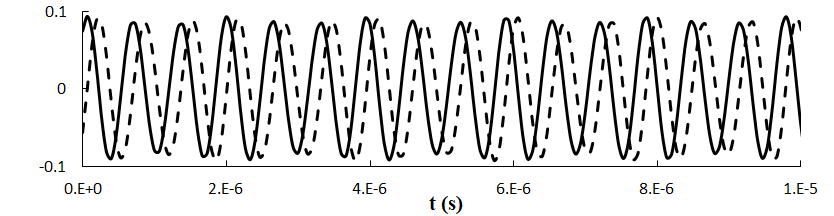}
      \caption{\rule[0.5ex]{0.5cm}{0.5pt} Model 5, \hdashrule[0.5ex]{0.5cm}{0.5pt}{1mm 0.5mm} Model 6}
      \label{fig:DSCFDCoef56}
    \end{subfigure}
    \quad
  \end{center}
  \caption{Temporal evolution of the POD mode amplitudes.}
  \label{fig:DStimeCoef}
\end{figure}

Another notable feature of the eigenvalue distribution is that the dominant POD modes appear in pairs with nearly identical energy levels. Figure~\ref{fig:DSmodes} presents the first six POD modes, while Fig.~\ref{fig:DStimeCoef} shows the temporal coefficients corresponding to these modes, obtained from the CFD results according to Eq.~(\ref{eq:rep}). Both figures indicate that each pair of modes exhibits similar spatial structures and identical frequencies, but are phase-shifted by approximately $\pi/2$ (in quadrature). This behavior is known due to translation symmetry associated with convective flow structures~\cite{holmes2010turbulence}. 

Furthermore, Fig.~\ref{fig:DStimeCoef} reveals that the mode pairs are harmonically related: the first and second modes correspond to the fundamental frequency, the third and fourth to the second harmonic, and the fifth and sixth to the third harmonic. From the flow-field perspective shown in Fig.~\ref{fig:DSmodes}, these harmonic relationships give rise to a transition from the dominant coherent structures to progressively finer-scale flow features. Meanwhile, the wavelength of the associated acoustic waves decreases correspondingly in the harmonic manner.

\begin{figure}
  \begin{center}
    \begin{subfigure}[t]{0.7\textwidth}
      \centering
      \includegraphics[width=0.9\textwidth]{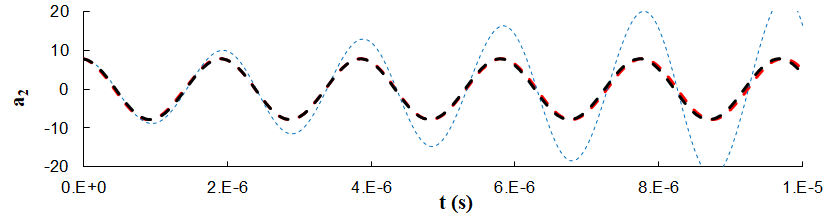}
      \caption{Mode 2}
      \label{fig:DSROMCoef2}
    \end{subfigure}
    \quad
    \begin{subfigure}[t]{0.7\textwidth}
      \centering
      \includegraphics[width=0.9\textwidth]{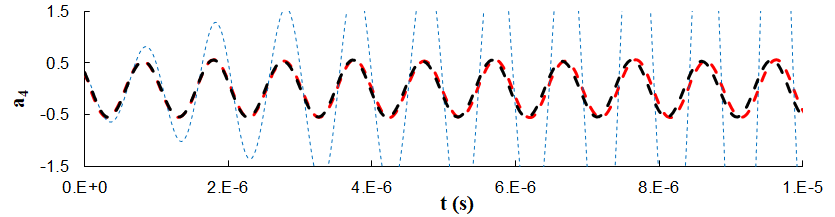}
      \caption{Mode 4}
      \label{fig:DSROMCoef4}
    \end{subfigure}
     \begin{subfigure}[t]{0.7\textwidth}
      \centering
      \includegraphics[width=0.9\textwidth]{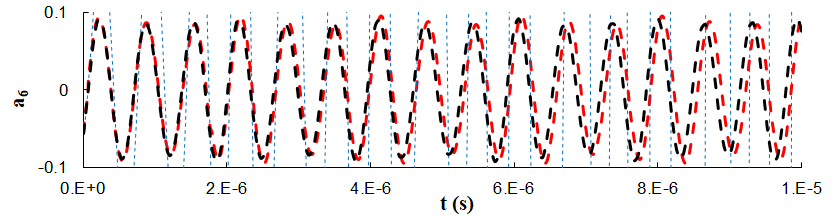}
      \caption{Mode 6}
      \label{fig:DSROMCoef6}
    \end{subfigure}
    \quad
  \end{center}
  \caption{Comparison of the temporal evolution of the POD mode amplitudes, \textcolor{black}{\hdashrule[0.5ex]{0.5cm}{0.5pt}{1mm 0.5mm}} CFD, \textcolor{blue}{\hdashrule[0.5ex]{0.5cm}{0.5pt}{1mm 0.5mm}} Galerkin-POD, and \textcolor{red}{\hdashrule[0.5ex]{0.5cm}{0.5pt}{1mm 0.5mm}} Bayesian-Galerkin-POD.}
  \label{fig:DSROMresults}
\end{figure}

While Fig.~\ref{fig:DStimeCoef} shows the dynamic behaviour of the modes extracted directly from the CFD results, the purpose of a ROM is to construct a system of ODEs that reproduces these dynamics. An accurate ROM should therefore be able to replicate the CFD results. The ODEs obtained from the standard Galerkin–POD formulation (Eqs.~(\ref{eq:rom}) and (\ref{eq:romops})) are compared with the CFD results in Fig.~\ref{fig:DSROMresults}. For clarity without loss of generality, only the even modes in each pair are shown here.

As seen in Fig.~\ref{fig:DSROMCoef2}, the standard Galerkin–POD (blue) shows good agreement with the CFD (black) in terms of both the dominant frequency and the oscillation amplitude during the first period. However, a well-known drawback of the standard Galerkin–POD is its instability: as time progresses, the discrepancy between the ROM and CFD solutions grows exponentially. By the fourth period, the deviation becomes comparable to the oscillation amplitude of the CFD data. Higher-order modes contain significantly less energy and are therefore more susceptible to ROM instability. For example, for the fourth mode shown in Fig.~\ref{fig:DSROMCoef4}, the discrepancy becomes comparable to the oscillation amplitude by the second period, while for the sixth mode in Fig.~\ref{fig:DSROMCoef6}, noticeable deviations appear even within the first period. 

\begin{figure}
  \begin{center}
    \begin{subfigure}[t]{0.32\textwidth}
      \centering
      \includegraphics[width=0.99\textwidth]{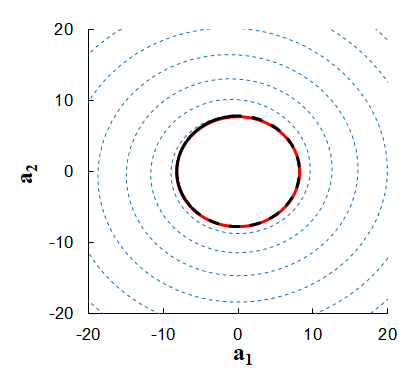}
      \caption{$a_{1}a_{2}$}
      \label{fig:DSROMDyna2}
    \end{subfigure}
    \begin{subfigure}[t]{0.32\textwidth}
      \centering
      \includegraphics[width=0.99\textwidth]{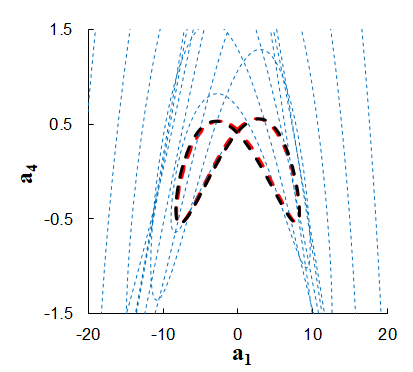}
      \caption{$a_{1}a_{4}$}
      \label{fig:DSROMDyna4}
    \end{subfigure}
     \begin{subfigure}[t]{0.32\textwidth}
      \centering
      \includegraphics[width=0.99\textwidth]{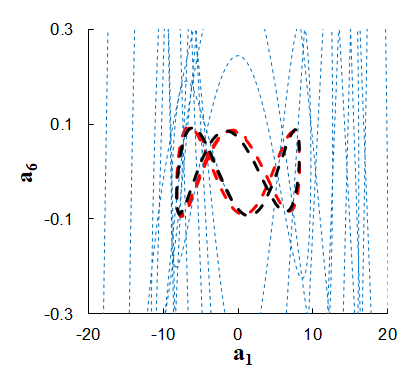}
      \caption{$a_{1}a_{6}$}
      \label{fig:DSROMDyna6}
    \end{subfigure}
  \end{center}
  \caption{Phase portraits (trajectories) of the POD modes, \textcolor{black}{\hdashrule[0.5ex]{0.5cm}{0.5pt}{1mm 0.5mm}} CFD, \textcolor{blue}{\hdashrule[0.5ex]{0.5cm}{0.5pt}{1mm 0.5mm}} Galerkin-POD, and \textcolor{red}{\hdashrule[0.5ex]{0.5cm}{0.5pt}{1mm 0.5mm}} Bayesian-Galerkin-POD}
  \label{fig:DSROMDyna}
\end{figure}

The corresponding dynamic patterns (phase portraits) of the first six modes in their respective phase spaces are shown in Fig.~\ref{fig:DSROMDyna}. From the CFD results, since modes 1 and 2 are in quadrature and possess comparable averaged energy, the dynamic relationship between $a^1$ and $a^2$ forms a limit cycle, represented as a closed circular trajectory. Similarly, the trajectories between $a^1-a^4$ and $a^1-a^6$ also exhibit closed-loop behaviours, reflecting the harmonic nature of the oscillations.

In contrast, the standard Galerkin–POD results show deviations due to the inherent instability of the method. The trajectory between $a^1$ and $a^2$ evolves into a spiral-type source around the CFD limit cycle, indicating growing amplitude with time. The closed-loop trajectories between $a^1-a^4$ and $a^1-a^6$ also diverge progressively, and the divergence becomes more pronounced over shorter timescales.

Since the first six modes account for $99.999\%$ of the total kinetic energy, the observed instability is not attributed to model uncertainty (i.e. POD mode truncation), but rather to data uncertainty. In fact, second-order velocity derivatives are required in Eq.~(\ref{eq:nsopermatrix}) and were post-processed in Tecplot. Extremely large values were observed near the wall and at vortex interfaces, indicating numerical noise and sensitivity in these regions. The present work therefore employs Bayesian inference Eq.~(\ref{eq:bimathmeanre})) to address such data-induced uncertainties, which are the primary cause of instability in the standard Galerkin–POD model.

When applying Bayesian inference, the prior probability distributions of the ODE system coefficients need to be specified. The expected values of distributions are naturally taken from the standard Galerkin–POD results obtained using Eqs.~(\ref{eq:rom}) and (\ref{eq:romops}). As no specific information is available regarding the deviation of these prior distributions, the standard deviations are simply assumed to be unity, and the coefficients are considered mutually independent. 

\begin{figure}
    \includegraphics[width=0.9\linewidth]{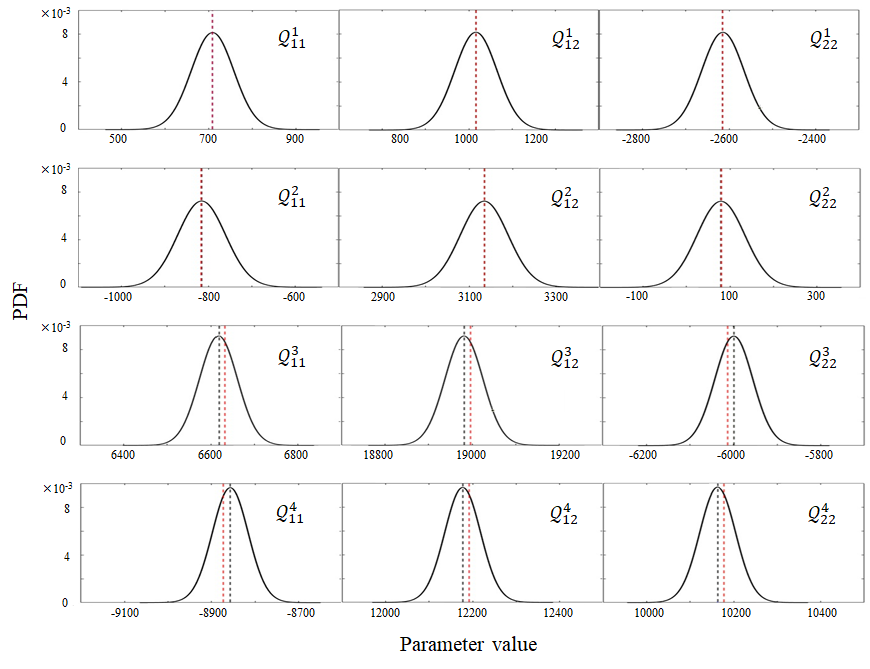}
    \caption{Posterior distribution of ROM coefficients.}
    \label{fig:DSQPDF}
\end{figure}

Figure~\ref{fig:DSQPDF} presents the results of Bayesian inference by showing the posterior probability distributions obtained from the statistical inverse problem. The first mode pair ($a^1-a^2$) accounts for approximately $99.5\%$ of the total energy; therefore, the coefficients representing the influence of this pair on the first four modes, namely $\mathcal{Q}_{11}^k$, $\mathcal{Q}_{12}^k$, and $\mathcal{Q}_{22}^k$ for $k=1-4$ in Eqs.~(\ref{eq:rom}), are shown in the figure (for simplicity, here the quadratic term coefficients $\mathcal{Q}_{1}$ and $\mathcal{Q}_{2}$ and combined into a single coefficient $\mathcal{Q}$). The expected values of both the posterior and prior probability distributions are also compared. The difference between the posterior and prior expectations represents a correction of the coefficient estimates, accounting for the uncertainties inherent in the standard Galerkin-POD.

As shown in Fig.~\ref{fig:DSQPDF}, for $k = 1, 2$, the prior prediction from the standard Galerkin–POD model is in excellent agreement with the Bayesian posterior expectation, indicating that the first two dominant modes require negligible correction. As discussed earlier, in this case the primary source of uncertainty arises from data post-processing in Tecplot rather than from POD mode truncation. The first $2$ modes correspond to the largest scale coherent structures of the flow, for which gradient-induced errors are minimal. As the modal index increases ($k = 3, 4$), finer-scale flow features emerge (illustrated in Fig.~\ref{fig:DSmodes}), leading to an accumulation of post-processing noise and a correspondingly larger degree of Bayesian correction (Fig.~\ref{fig:DSQPDF}). Nevertheless, even for $k = 3, 4$, while the absolute magnitudes of the coefficients are on the order of $10^3$–$10^4$, the corrections are only around $10$. This demonstrates that the standard Galerkin–POD already captures the flow dynamics with satisfactory accuracy. However, even small deviations in the reduced coefficients can induce significant instability over time. These findings highlight the pronounced sensitivity of Galerkin-based reduced-order models to uncertainties, reinforcing the need for statistical correction mechanisms to ensure predictive reliability.

The ODE system constructed using the Bayesian inference corrected coefficients is referred to as the Bayesian–Galerkin–POD model in this study. Figure~\ref{fig:DSROMCoef2} illustrates the impact of these corrections on ROM accuracy and stability. Although the variance of the prior coefficient distributions was simply assumed to be unity, the Bayesian–Galerkin–POD shows a substantial improvement over the standard Galerkin–POD, achieving excellent agreement with the original CFD results. In particular, for the sixth mode, the amplitude of the successive oscillation is characterised by a high–low–low sequence. This complicated behavior is also accurately reproduced by the Bayesian–Galerkin–POD. Consequently, as shown in Fig.~\ref{fig:DSROMDyna}, the corresponding phase portraits exhibit nearly identical dynamic trajectories to those obtained from the CFD data.

\begin{figure}
  \begin{center}
    \begin{subfigure}[t]{0.4\textwidth}
      \centering
      \includegraphics[width=0.9\textwidth]{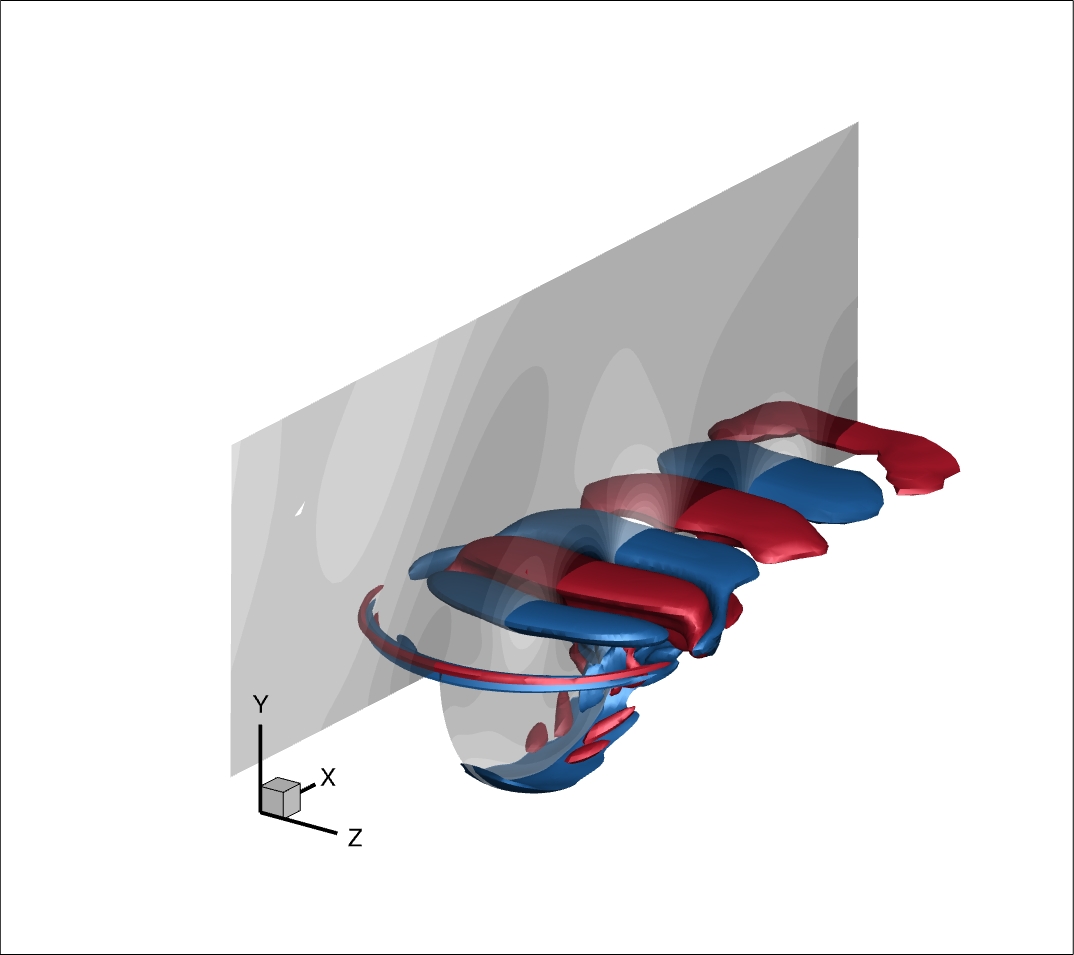}
      \caption{$s=0$}
      \label{fig:DSReFF1}
    \end{subfigure}
    \quad
    \begin{subfigure}[t]{0.4\textwidth}
      \centering
      \includegraphics[width=0.9\textwidth]{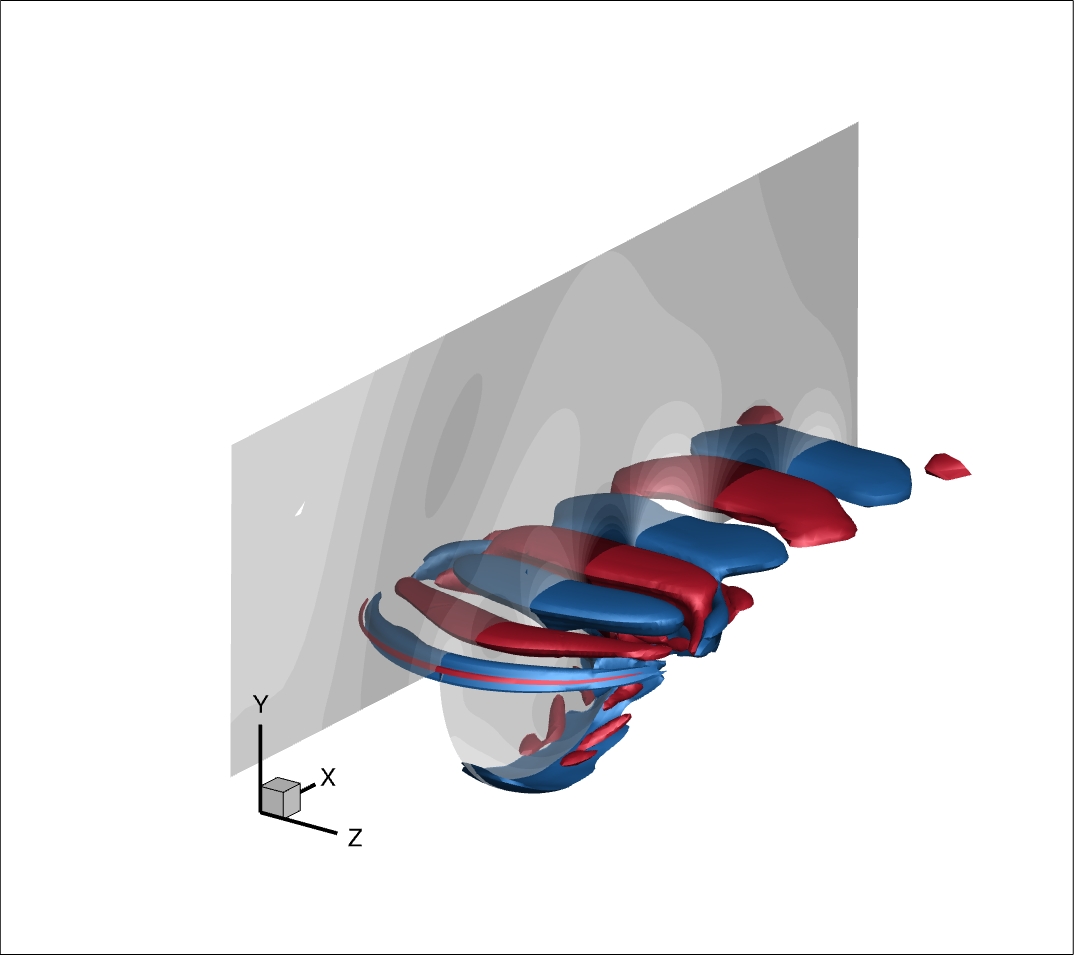}
      \caption{$s=0.25$}
      \label{fig:DSReFF2}
    \end{subfigure}
     \begin{subfigure}[t]{0.4\textwidth}
      \centering
      \includegraphics[width=0.9\textwidth]{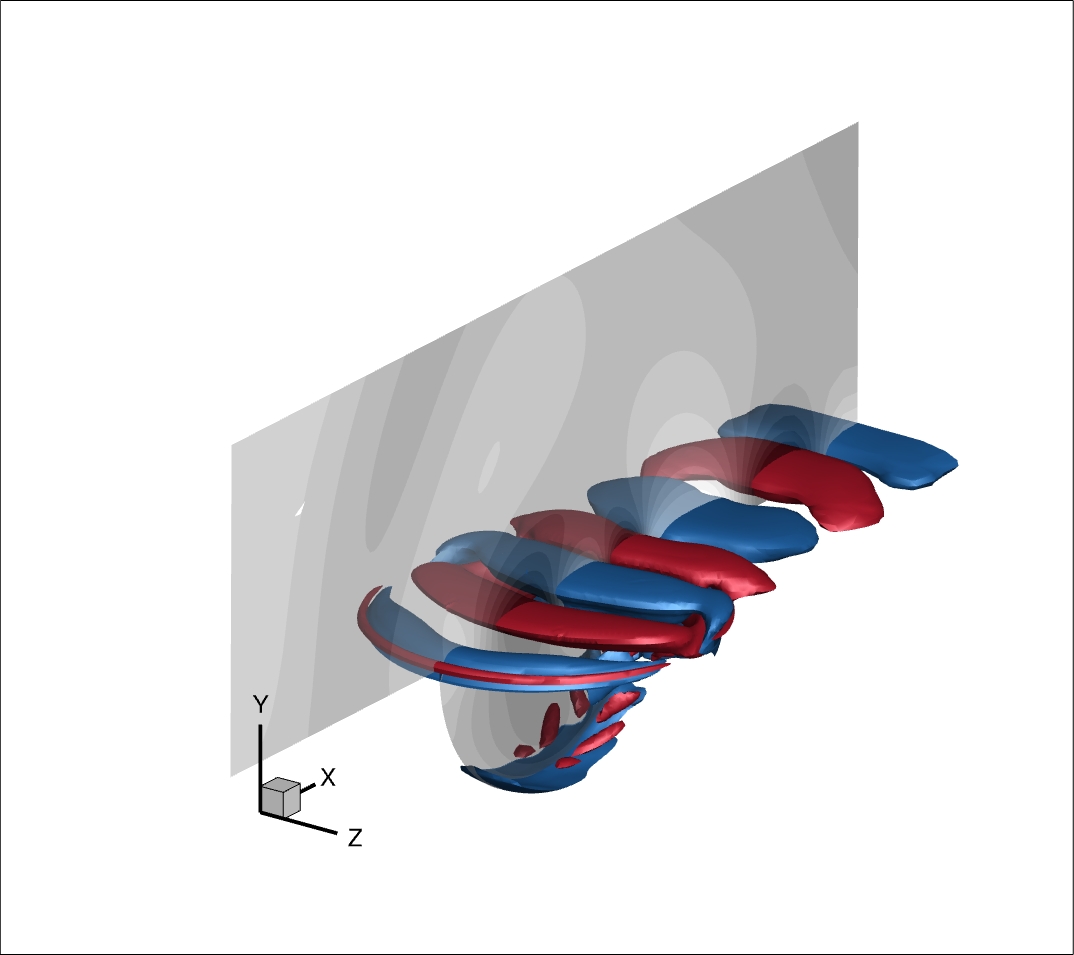}
      \caption{$s=0.5$}
      \label{fig:DSReFF3}
    \end{subfigure}
    \quad
    \begin{subfigure}[t]{0.4\textwidth}
      \centering
      \includegraphics[width=0.9\textwidth]{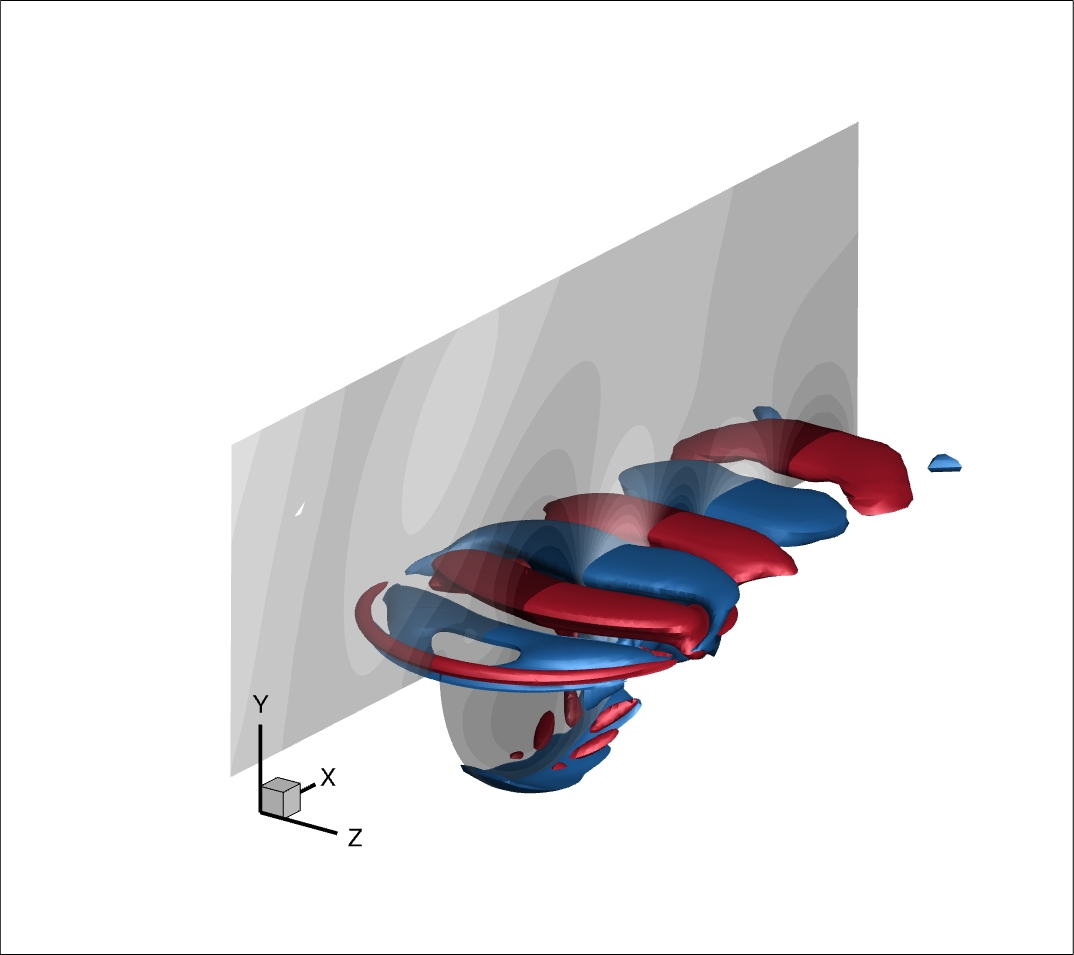}
      \caption{$s=0.75$}
      \label{fig:DSReFF4}
    \end{subfigure}
  \end{center}
  \caption{Flow field reconstruction based on Bayesian-Galerkin-POD}
  \label{fig:DSReFF}
\end{figure}

Finally, the three-dimensional unsteady flow field was reconstructed using the Bayesian enhanced Galerkin–POD model based on the first six modes. The reconstructed flow field over one fundamental harmonic period is shown in Fig.~\ref{fig:DSReFF}. As mentioned earlier, the ROM was performed on a smaller computational domain with a uniform but coarser mesh; therefore, the visualised region is correspondingly smaller. Apart from the difference in domain size, there is almost no visible discrepancy between Fig.~\ref{fig:DSReFF} and the original CFD results shown in Fig.~\ref{fig:DSFlowfield}, further validating the accuracy and reliability of the proposed ROM framework.

\section{\textbf{Centrifugal Compressor}}

\subsection{\textbf{Physical problem and CFD results}}

To demonstrate the capability of the present framework in addressing practical problems with high Reynolds numbers and complex geometries, the Bayesian–Galerkin–POD method is applied to a centrifugal compressor. This case is considerably more challenging than the dimpled surface flow. First, it involves a high-Reynolds-number ($Re \approx 10^5$) turbulent flow, where the energy spectrum is smoother and this makes modal truncation difficult. Second, unlike the dimpled surface case that features a single dominant flow structure (the shear layer), multiple flow phenomena (such as tip-leakage flow, impeller–diffuser interaction) are strongly coupled, leading to highly complex unsteady dynamics.

The configuration of the compressor is shown in Fig.~\ref{fig:ComConfig}. A centrifugal compressor primarily consists of a rotating impeller and a stationary diffuser. The impeller transfers mechanical energy to the working fluid: as a fluid particle passes through the impeller, it gains energy through an increase in absolute tangential velocity. Consequently, both the total pressure and total temperature of the flow rise. The diffuser further increases the static pressure by expanding the cross-sectional area and redirecting the flow through its vane-shaped passages. 
\begin{figure}
  \begin{center}
    \begin{subfigure}[t]{0.45\textwidth}
      \centering
      \includegraphics[width=0.9\textwidth]{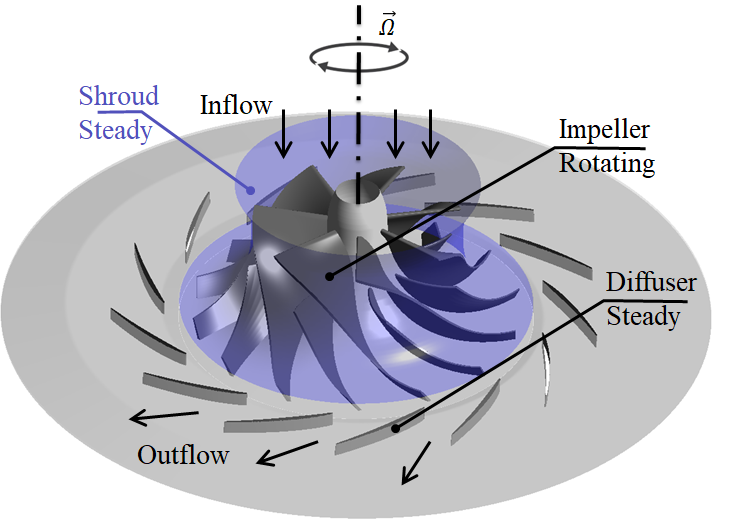}
      \caption{3D view}
      \label{fig:ComConfig}
    \end{subfigure}
    \quad
    \begin{subfigure}[t]{0.44\textwidth}
      \centering
      \includegraphics[width=0.9\textwidth]{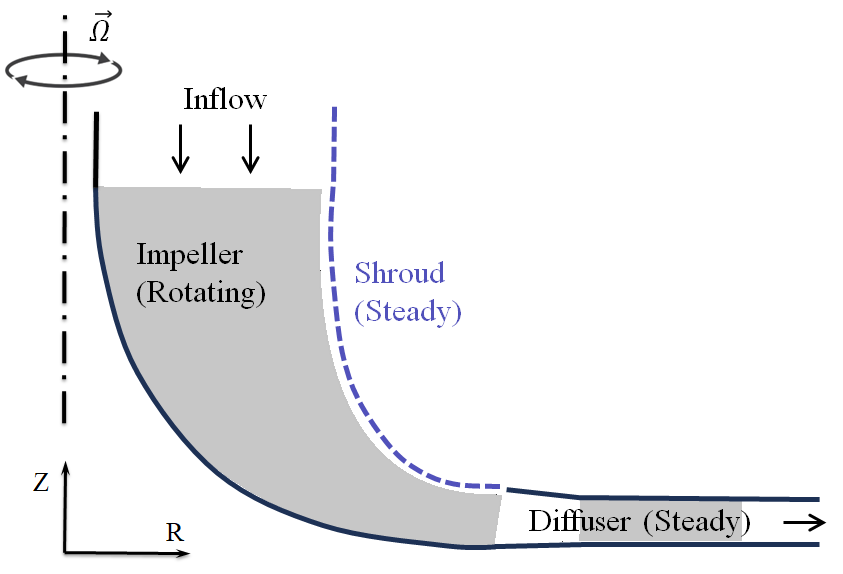}
      \caption{2D Z-R plane view}
      \label{fig:Comconfig2}
    \end{subfigure}
  \end{center}
  \caption{Centrifugal compressor configuration.}
  \label{fig:ComConfig}
\end{figure}

The geometry and operating conditions of the compressor used in the present study are listed in Table~\ref{table:ComGeom}. The high rotational speed gives rise to significant flow-field unsteadiness between the rotating and stationary components. The impeller–shroud interaction gives rise to the well-known tip-leakage vortex: the rotating impeller is enclosed by a stationary casing (shroud), and a small clearance (approximately $3\%$ of the blade radius) exists between them. This clearance acts similarly to the tip of an aircraft wing, across which a significant pressure difference develops. The resulting pressure difference generates a tornado-like tip-leakage flow, which subsequently breaks down within the passage and interacts with the other blades. The impeller–diffuser interaction leads to the well-known stage-to-stage unsteadiness: the impeller outlet flow contains strong wakes, and as each impeller blade passes a diffuser vane, the vane experiences periodically varying inlet flow conditions due to these impeller wakes.
\begin{table}[t]
\caption{Geometry and operation conditions of the studied compressor.}
\begin{center}
\begin{tabular}{c c}
\hline
    Parameter	&Value \\
\hline
Impeller blade number  & $7+7$ (main and splitter)	 \\
Diffuser vane number  & $13$ \\
Impeller exit diameter (D)  & $102 ~mm$	 \\
Rotation speed ($\Omega$) & $9000 ~rad/s $	 \\
Reynold number ($\frac{\Omega D^2 \rho}{\mu}$)  & $\sim 1.8\times 10^5 $	 \\
Pressure ratio  & $\sim 2.1 $	 \\
\hline
\end{tabular}
	\label{table:ComGeom}
\end{center}
\end{table}

\begin{figure}
    \includegraphics[width=0.5\linewidth]{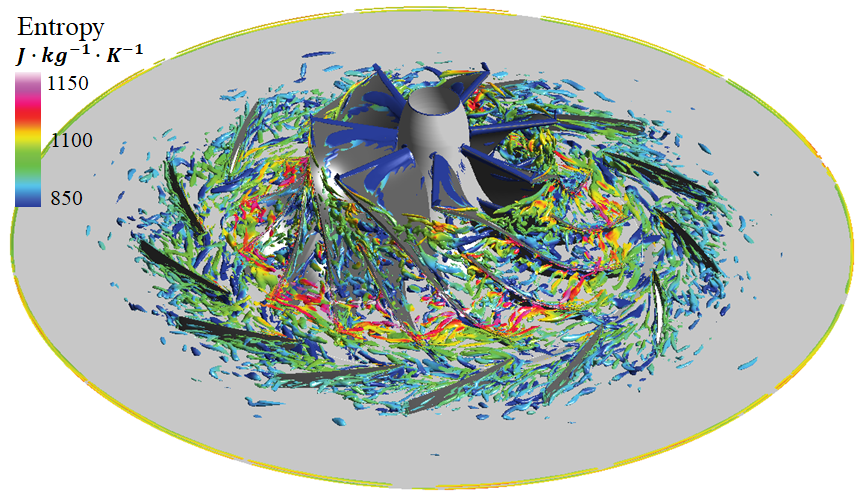}
    \caption{Q criterion of tip-leakage vortex and impeller-diffuser interactions.}
    \label{fig:ComLamda2}
\end{figure}

Figure~\ref{fig:ComLamda2} presents iso-surfaces of the Q-criterion to visualise the flow-field unsteadiness within the compressor, clearly revealing the tip-leakage flow and the impeller–diffuser interaction. The contours in the figure represent entropy, where higher values correspond to regions of intense local turbulent dissipation. The results are obtained from Large Eddy Simulation (LES) using the Smagorinsky subgrid-scale model. All solid boundaries are treated as adiabatic and no-slip. The computational domain extends ten impeller diameters upstream of the impeller inlet to ensure fully developed inflow conditions. Structured meshes are employed for both the impeller and diffuser passages, with near-wall refinement corresponding to 
$y^+  \approx1$,  $x^+  \approx 50$ in the streamwise direction, and $z^+  \approx25$ in the spanwise direction. The simulations are carried out using ANSYS CFX and covered fifty full impeller rotation periods. The time step is chosen such that the impeller rotates by $0.5^o$ per step. The reuslts of the last two rotation periods are stored every two time steps. Hence, in total, $720$ snapshots are saved, each sanpshot with a data size of approximately $10$ GB.

\subsection{\textbf{Application of Bayesian-Galerkin-POD}}

\begin{figure}
    \includegraphics[width=0.3\linewidth]{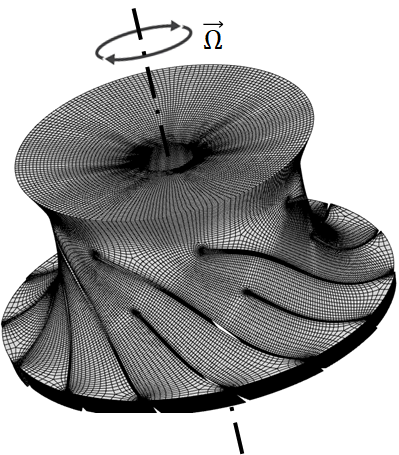}
    \caption{Course and more uniform mesh for results interpolation.}
    \label{fig:ComCorseMesh}
\end{figure}

\begin{figure}
    \includegraphics[width=0.4\linewidth]{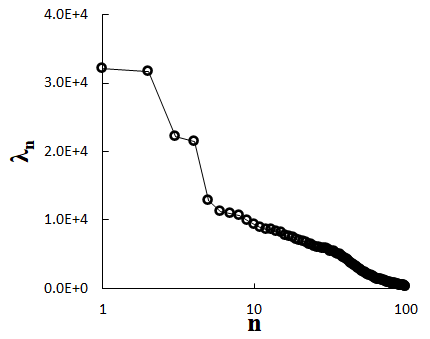}
    \caption{POD eigenvalues.}
    \label{fig:ComLMD}
\end{figure}

Similar to the dimpled surface case, the CFD results were interpolated onto a coarser and more uniform mesh prior to performing the POD (as illustrated in Fig.~\ref{fig:ComCorseMesh}) to reduce the size of the SVD decomposition matrix. The eigenvalues of the POD modes, with the mean flow removed, are presented in Fig.~\ref{fig:ComLMD}. A comparison between Fig.~\ref{fig:DSFlowfield} and Fig.~\ref{fig:ComLMD} reveals a notable difference: for the centrifugal compressor case, the modal energy decays much more slowly. Even for the twentith modes, the corresponding eigenvalue remains approximately $25\%$ of the first two modes. This characteristic complicates POD mode truncation and highlights the strong coupling between flow structures of different scales.

\begin{figure}
  \begin{center}
    \begin{subfigure}[t]{0.3\textwidth}
      \centering
      \includegraphics[width=0.99\textwidth]{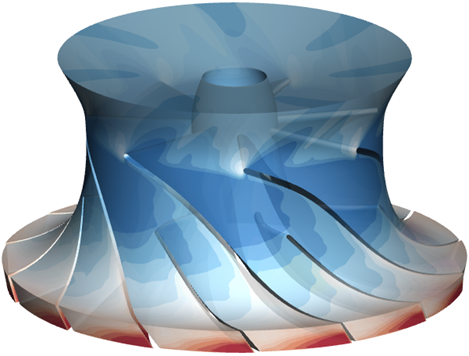}
      \caption{Mean flow}
      \label{fig:ROMMOD0}
    \end{subfigure}
    \begin{subfigure}[t]{0.3\textwidth}
      \centering
      \includegraphics[width=0.99\textwidth]{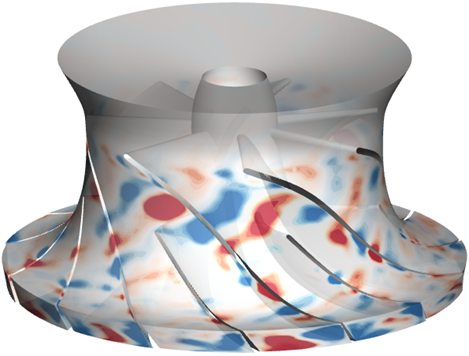}
      \caption{Mode 1}
      \label{fig:ROMMOD1}
    \end{subfigure}
     \begin{subfigure}[t]{0.3\textwidth}
      \centering
      \includegraphics[width=0.99\textwidth]{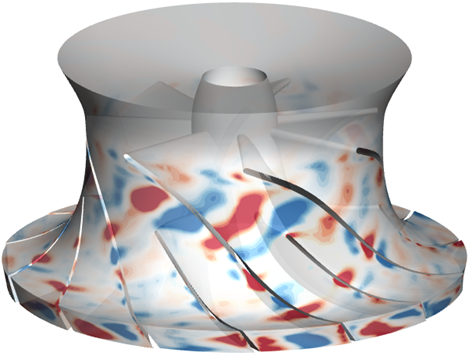}
      \caption{Mode 2}
      \label{fig:ROMMOD2}
    \end{subfigure}
        \begin{subfigure}[t]{0.3\textwidth}
      \centering
      \includegraphics[width=0.99\textwidth]{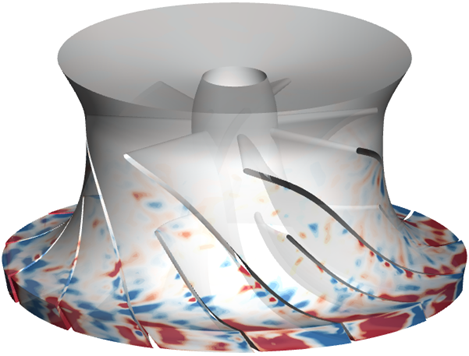}
      \caption{Mode 3}
      \label{fig:ROMMOD3}
    \end{subfigure}
     \begin{subfigure}[t]{0.3\textwidth}
      \centering
      \includegraphics[width=0.99\textwidth]{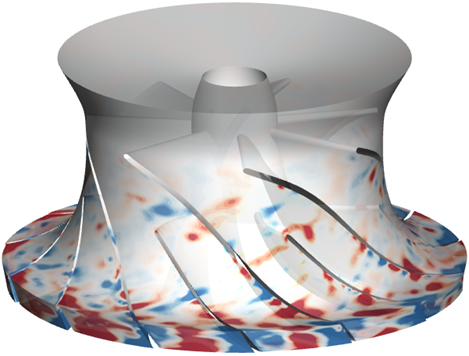}
      \caption{Mode 4}
      \label{fig:ROMMOD4}
    \end{subfigure}
  \end{center}
  \caption{Mean flow and first four POD modes.}
  \label{fig:ComROMMOD}
\end{figure}

Seen from Fig.~\ref{fig:ComLMD}, the first four modes show similarity to the dimple surface case: they exhibit obviously higher energy levels and appear in pairs. These four modes, together with the mean flow field, are illustrated in Fig.~\ref{fig:ComROMMOD}, where contours of density are shown. The features of the first two modes originate from the leading edge of the main blade and extend into the blade passage, clearly revealing a vortex breakdown pattern. This indicates that the first two modes correspond to the breakdown of the tip-leakage flow. In contrast, the features of the third and fourth modes are primarily located near the impeller outlet, implying that these modes are mainly associated with the impeller–diffuser interaction.

\begin{figure}
  \begin{center}
    \begin{subfigure}[t]{0.7\textwidth}
      \centering
      \includegraphics[width=0.95\textwidth]{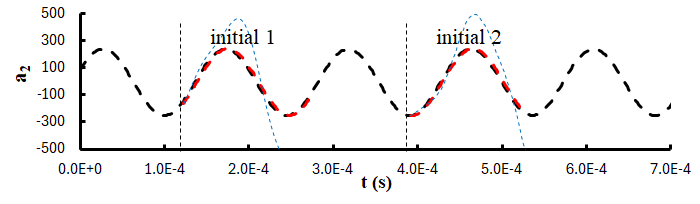}
      \caption{Mode 2}
      \label{fig:Coma1}
    \end{subfigure}
    \quad
    \begin{subfigure}[t]{0.7\textwidth}
      \centering
      \includegraphics[width=0.95\textwidth]{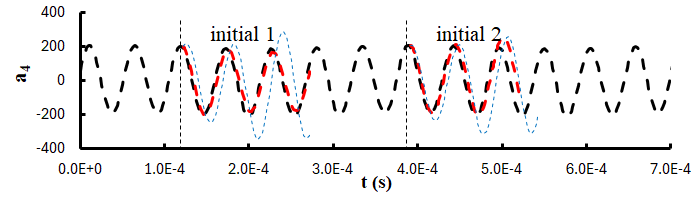}
      \caption{Mode 4}
      \label{fig:Coma1}
    \end{subfigure}
    \quad
    \begin{subfigure}[t]{0.7\textwidth}
      \centering
      \includegraphics[width=0.95\textwidth]{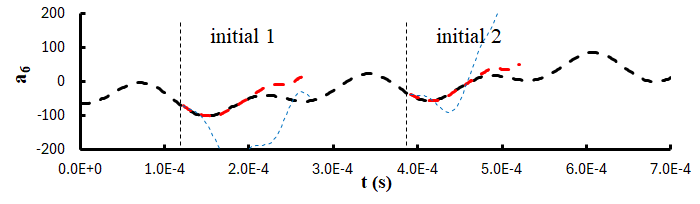}
      \caption{Mode 6}
      \label{fig:Coma1}
    \end{subfigure}
    \quad
    \begin{subfigure}[t]{0.7\textwidth}
      \centering
      \includegraphics[width=0.95\textwidth]{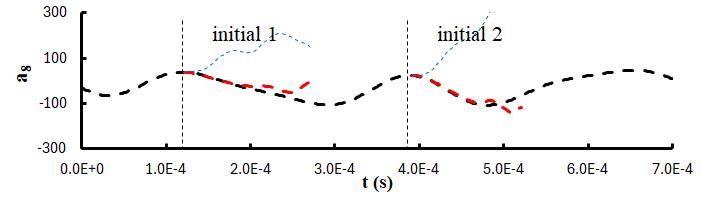}
      \caption{Mode 8}
      \label{fig:Coma1}
    \end{subfigure}
    \quad
     \begin{subfigure}[t]{0.7\textwidth}
      \centering
      \includegraphics[width=0.95\textwidth]{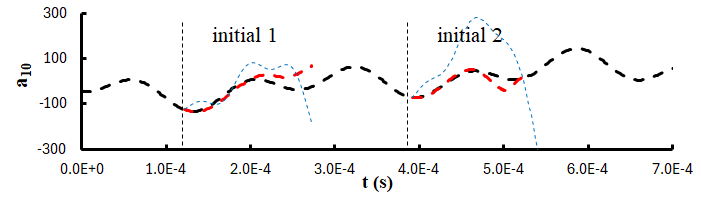}
      \caption{Mode 10}
      \label{fig:Coma1}
    \end{subfigure}
  \end{center}
   \caption{Comparison of the temporal evolution of the POD mode amplitudes, \textcolor{black}{\hdashrule[0.5ex]{0.5cm}{0.5pt}{1mm 0.5mm}} CFD, \textcolor{blue}{\hdashrule[0.5ex]{0.5cm}{0.5pt}{1mm 0.5mm}} Galerkin-POD, and \textcolor{red}{\hdashrule[0.5ex]{0.5cm}{0.5pt}{1mm 0.5mm}} Bayesian-Galerkin-POD.}
  \label{fig:COMROMresults}
\end{figure}

The relationship between these four modes and the flow-field unsteadiness, specifically the tip-leakage vortex and impeller–diffuser interactions, can be further evaluated through the temporal evolution of the POD mode coefficients. Figure~\ref{fig:COMROMresults} presents the time variation of the even-numbered mode coefficients, from Mode 2 to Mode 10, over one impeller rotation period. It can be observed that the oscillation frequency of $a_2$ is approximately five times the rotational frequency. Since $a_2$ corresponds to the tip-leakage vortex breakdown, this frequency represents the characteristic vortex breakdown frequency. In contrast, the oscillation frequencies of $a_4$ is about thirteen times the rotational frequency. This matches the number of diffuser vanes, hence indicates that the fourth POD mode capture the unsteady interaction between the rotating impeller and the stationary diffuser.

As the modal energy decays slowly, approximately 46 modes are required to capture $80\%$ of the total unsteady energy in this case. Since the reduced-order model (ROM) includes second-order terms of the form $a_ia_j$, employing $46$ modes results in a coefficient matrix $A$ (as described by Eq.~(\ref{eq:rom})) of size $46\times 1151$. Such a large system is impractical for an efficient ROM. Therefore, in the present study, only the first ten modes are retained to construct the ROM, which captures around $40\%$ of the overall unsteady flow energy ($\sum_{i=1}^{10} \lambda_{i} / \sum_{i=1}^{700} \lambda_{i}=0.378$). The effects of the neglected higher-order modes are implicitly treated as uncertainties, which are accounted for through the Bayesian inference framework. As around $60\%$ of the overall unsteady energy has been truncated, the main uncertainty of this case is model uncertainty rather than data uncertainty.

Since roughly $60\%$ of the unsteady energy has been truncated, the dominant source of uncertainty in this case arises from significant model reduction rather than data noise. Furthermore, this configuration involves multiple interacting flow features, specifically the tip-leakage flow and the impeller–diffuser interaction, with each characterised by distinct dominant frequencies. Under such challenging conditions, the Bayesian–Galerkin–POD framework continues to exhibit strong reliability and robustness. Figure~\ref{fig:COMROMresults} compares the performance of the classical Galerkin–POD and the Bayesian–Galerkin–POD, using LES results as the reference over one full impeller rotation period. Both reduced-order models are initialised from two distinct flow states and integrated over approximately one-fifth of an impeller rotation period, corresponding to the characteristic timescale of the tip-leakage vortex breakdown. As shown in the figure, the Bayesian inference substantially enhances the predictive accuracy of the conventional Galerkin–POD.

For modes $a_2$ and $a_4$ (correspond to the modal pairs $a_1$–$a_2$ and $a_3$–$a_4$ respectively), the oscillation frequencies are correctly captured by the classical Galerkin–POD. However, noticeable amplitude deviations emerge almost immediately. Even for the dominant mode $a_2$, the prediction diverges within a single tip-leakage vortex period, indicating limited stability of the standard ROM. For the higher modes ($a_6$–$a_{10}$), the dynamics become significantly more intricate, and the standard Galerkin–POD remains reliable only within a narrow temporal window of approximately $\Delta t \approx 1\times10^{-5}$ s. In contrast, the Bayesian–Galerkin–POD provides a robust reconstruction of the unsteady flow dynamics across all examined modes. Even for the last three modes, the results remain in excellent agreement with the CFD benchmark up to $\Delta t \approx 1\times10^{-4}$ s. This means the Bayesian-Galerkin-POD has extended the model’s predictable time window by nearly an order of magnitude compared to the classical Galerkin–POD.

\begin{figure}
  \begin{center}
    \begin{subfigure}[t]{0.3\textwidth}
      \centering
      \includegraphics[width=0.99\textwidth]{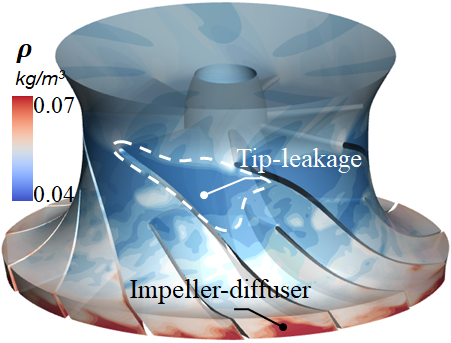}
      \caption{$t=1.2\times 10^{-5}s$}
      \label{fig:ComCFD1}
    \end{subfigure}
    \begin{subfigure}[t]{0.3\textwidth}
      \centering
      \includegraphics[width=0.99\textwidth]{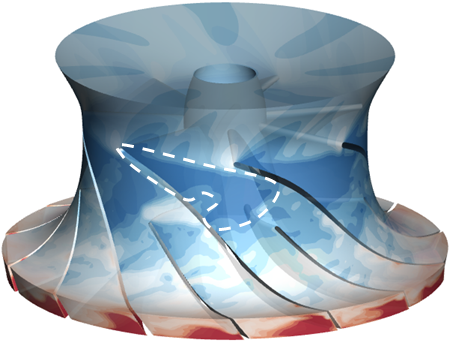}
      \caption{$t=4.7\times 10^{-5}s$}
      \label{fig:ComCFD2}
    \end{subfigure}
     \begin{subfigure}[t]{0.3\textwidth}
      \centering
      \includegraphics[width=0.99\textwidth]{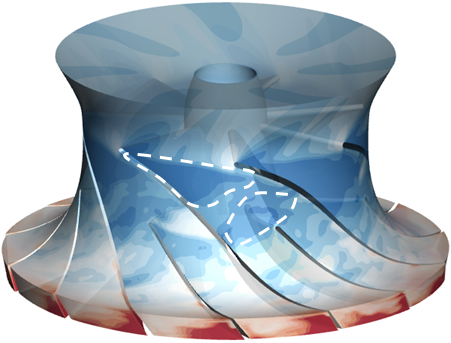}
      \caption{$t=8.2\times 10^{-5}s$}
      \label{fig:ComCFD3}
    \end{subfigure}
  \end{center}
  \caption{3D unsteady flow field from LES}
  \label{fig:ComLESTL}
\end{figure}

\begin{figure}
  \begin{center}
    \begin{subfigure}[t]{0.3\textwidth}
      \centering
      \includegraphics[width=0.99\textwidth]{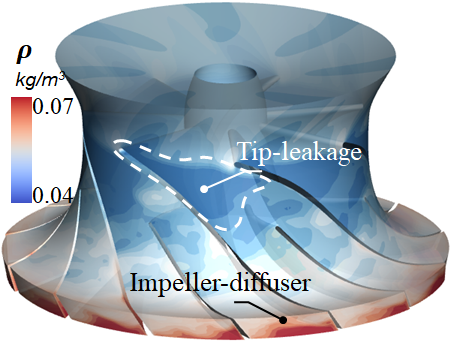}
      \caption{$t=1.2\times 10^{-5}s$}
      \label{fig:ComROM1}
    \end{subfigure}
    \begin{subfigure}[t]{0.3\textwidth}
      \centering
      \includegraphics[width=0.99\textwidth]{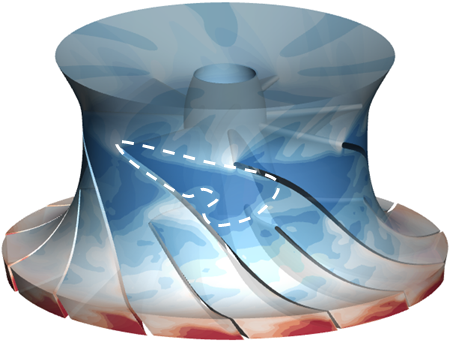}
      \caption{$t=4.7\times 10^{-5}s$}
      \label{fig:ComROM2}
    \end{subfigure}
     \begin{subfigure}[t]{0.3\textwidth}
      \centering
      \includegraphics[width=0.99\textwidth]{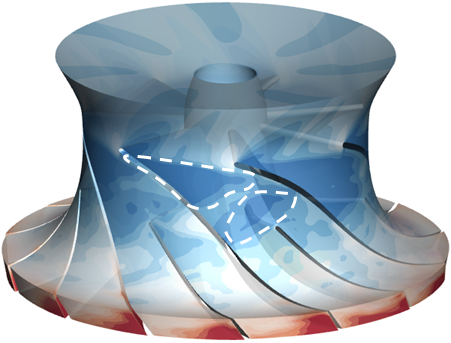}
      \caption{$t=8.7\times 10^{-5}s$}
      \label{fig:ComROM3}
    \end{subfigure}
  \end{center}
  \caption{Reconstructed 3D unsteady flow field based on Bayesian-Galerkin-POD using POD 10 modes}
  \label{fig:ComROMTL}
\end{figure}

Finally, the three-dimensional LES results are compared with the Bayesian–Galerkin–POD predictions in Figure~\ref{fig:ComLESTL} and Figure~\ref{fig:ComROMTL}. Both simulations are initialised from the first condition shown in Figure~\ref{fig:COMROMresults} and integrated over approximately one tip-leakage vortex breakdown period. As illustrated, both LES and ROM successfully capture the vortex breakdown process induced by the tip-leakage flow. In addition, the evolution of the flow field at the impeller exit (impeller-diffuser interaction) has also been well captured. While the LES provides highly detailed flow-field information and resolves the fine-scale turbulent structures, the reduced-order model reproduces the essential large-scale dynamics with excellent qualitative and quantitative agreement. Remarkably, the ROM achieves this predictive accuracy at a negligible computational cost compared to the LES, highlighting the capability and efficiency of the proposed Bayesian–Galerkin–POD framework for complex flows.

\section{Conclusion}

This work has presented a novel Bayesian-enhanced Galerkin–POD framework for unsteady compressible flows, designed to overcome the fundamental instability and limited accuracy of conventional reduced-order models (ROMs). By formulating the coefficient calibration of the Galerkin-projected ODE system as a statistical inverse problem, the proposed method explicitly accounts for both model uncertainty, arising from truncation of low-energy modes, and data uncertainty, introduced through measurement noise and numerical differentiation. Bayesian inference provides a systematic and probabilistic means of refining the ROM coefficients, where prior knowledge derived from the Galerkin projection is updated through the observed temporal coefficients of the POD modes. The analytical solution based on Gaussian likelihood and inverse-Gamma priors allows efficient implementation without the need for costly sampling algorithms, making the approach suitable for high-dimensional turbulent systems.

The performance of the proposed framework has been demonstrated using two representative compressible flow configurations of increasing complexity. The first is a moderate-Reynolds-number case ($Re = 3000$) featuring a single dominant flow structure: a self-sustained oscillating shear layer over a dimpled surface. The second is a high-Reynolds-number case ($Re \approx 1.8\times10^5$) involving multiple interacting flow features in a centrifugal compressor, characterised by strong tip-leakage flow and impeller–diffuser interactions. In the dimpled-surface case, the Bayesian–Galerkin–POD model accurately reproduced the oscillatory behaviour, modal amplitudes, and phase-space trajectories obtained from DNS, while eliminating the divergence typically observed in the standard Galerkin–POD. In the turbulent compressor case, the framework maintained robust performance despite retaining only a limited number of POD modes, successfully capturing dominant unsteady phenomena such as tip-leakage vortex breakdown and rotating–stationary coupling. Collectively, these results demonstrate that the proposed method substantially enhances the stability and predictive capability of reduced-order models across a wide range of flow regimes.

Overall, the present framework combines the physical interpretability of Galerkin–POD with the robustness of Bayesian inference, providing a reliable and computationally efficient approach for modelling complex unsteady compressible flows. Future work will explore its integration into digital-twin architectures for real-time flow prediction and control.


\nocite{*}

\bibliography{apssamp}

\end{document}